\documentclass{IEEEtran4PSCC}
\ifCLASSINFOpdf
   \usepackage[pdftex]{graphicx}
\else
   \usepackage[dvips]{graphicx}
\fi
\usepackage{enumitem}
\usepackage{dblfloatfix}

\usepackage[cmex10]{amsmath}
\usepackage{amssymb}
\usepackage{algorithm}
\usepackage[noend]{algpseudocode}

\usepackage{subcaption}
\usepackage{graphicx}

\usepackage{enumitem}

\usepackage{xcolor}

\usepackage{subfiles}
\usepackage{tikz}
\usepackage{svg}
\usepackage{hyperref}

\usepackage{tablefootnote}

\usepackage{globalvals}

\usepackage{bbold}

\makeatletter
\let\old@ps@headings\ps@headings
\let\old@ps@IEEEtitlepagestyle\ps@IEEEtitlepagestyle
\def\psccfooter#1{%
    \def\ps@headings{%
        \old@ps@headings%
        \def\@oddfoot{\strut\hfill#1\hfill\strut}%
        \def\@evenfoot{\strut\hfill#1\hfill\strut}%
    }%
    \def\ps@IEEEtitlepagestyle{%
        \old@ps@IEEEtitlepagestyle%
        \def\@oddfoot{\strut\hfill#1\hfill\strut}%
        \def\@evenfoot{\strut\hfill#1\hfill\strut}%
    }%
    \ps@headings%
}
\makeatother

\begin{document}
\title{Self-Supervised Graph Neural Networks\\for Full-Scale Tertiary Voltage Control}
\author{
\IEEEauthorblockN{B. Donon, G. Jamgotchian, H. Kulesza}
\IEEEauthorblockA{Department of R\&D\\
RTE (Réseau de Transport d'Électricité)\\
Paris, France}
\and
\IEEEauthorblockN{L. Wehenkel}
\IEEEauthorblockA{Department of EE \& CS \\
Université de Liège\\
Liège, Belgium}
}
\maketitle

\begin{abstract}

A growing portion of operators' workload is dedicated to \emph{Tertiary Voltage Control} (TVC), namely the regulation of voltages by means of adjusting a series of setpoints and connection status.
TVC may be framed as a \emph{Mixed Integer Non Linear Program}, but state-of-the-art optimization methods scale poorly to large systems, making them impractical for real-scale and real-time decision support.
Observing that TVC does not require any optimality guarantee, we frame it as an \emph{Amortized Optimization} problem, addressed by the self-supervised training of a \emph{Graph Neural Network} (GNN) to minimize voltage violations.
As a first step, we consider the specific use case of post-processing the forecasting pipeline used by the French TSO, where the trained GNN would serve as a TVC proxy.
After being trained on one year of full-scale HV-EHV French power grid 
day-ahead forecasts, our model manages to significantly reduce the average number of voltage violations. 

\end{abstract}



\section{Background \& Motivations}
\label{sec:background}

In addition to monitoring active power flows throughout the grid, operators are in charge of regulating voltage magnitudes to keep them within an acceptable range.
Too low or too high voltages can lead to voltage instability, \textit{i.e.} a black-out.
To regulate voltages, operators may adjust the voltage setpoints of \textit{Secondary Voltage Regulations}\footnote{Closed-loop regulation that couples the reactive power of multiple generators to regulate the voltage of a single regulated bus.} (SVRs) and of \textit{Ratio Tap Changers} (RTCs), and choose to connect or disconnect a series of pre-identified shunts, lines and synchronous condensers.
Until recently, this complex real-time decision making problem -- referred to as \emph{Tertiary Voltage Control} (TVC) -- only took a marginal part of the operators' daily routine.
However, ongoing changes occurring in the European energy ecosystem -- namely the growing share of renewable energies which reduce active power flows and induce reactive power, the burying of transmission lines, and new market mechanisms -- have drastically increased the rate and intensity of high voltage events that require an operator's decision.
We aim at developing a real-time automatic decision support tool to assist operators in this vital and complex task.

\begin{table*}[b!]
    \centering
    \begin{tabular}{|c l|l|l|l|}
        \hline
         & Class & Ports ($x$) & Context Features ($x$) & Decision Features ($y$) \\
        \hline
        \resizebox{0.04\linewidth}{!}{\tikzset{every picture/.style={line width=0.75pt}} 

\begin{tikzpicture}[x=0.75pt,y=0.75pt,yscale=-1,xscale=1]

\draw [color={rgb, 255:red, 0; green, 0; blue, 0 }  ,draw opacity=1 ][line width=3]    (25,20) -- (25,10) ;
\draw [line width=1.5]    (25,15) -- (35,15) ;
\draw  [draw opacity=0][fill={rgb, 255:red, 0; green, 0; blue, 0 }  ,fill opacity=1 ] (32,15) .. controls (32,13.34) and (33.34,12) .. (35,12) .. controls (36.66,12) and (38,13.34) .. (38,15) .. controls (38,16.66) and (36.66,18) .. (35,18) .. controls (33.34,18) and (32,16.66) .. (32,15) -- cycle ;
\draw  [draw opacity=0] (10,10) -- (40,10) -- (40,20) -- (10,20) -- cycle ;

\end{tikzpicture}} & Bus & Bus & $V$, $\vartheta$, $V_\text{nom}$, $\overline{V}$, $\underline{V}$, $\mathbb{1}_\text{opt}$ & - \\
        \resizebox{0.04\linewidth}{!}{\tikzset{every picture/.style={line width=0.75pt}} 

\begin{tikzpicture}[x=0.75pt,y=0.75pt,yscale=-1,xscale=1]

\draw [line width=1.5]    (30,15) -- (35,15) ;
\draw  [line width=1.5]  (20,10) -- (30,10) -- (30,20) -- (20,20) -- cycle ;
\draw [line width=1.5]    (20,10) -- (30,20) ;
\draw [line width=1.5]    (20,20) -- (30,10) ;

\draw  [draw opacity=0][fill={rgb, 255:red, 0; green, 0; blue, 0 }  ,fill opacity=1 ] (32,15) .. controls (32,13.34) and (33.34,12) .. (35,12) .. controls (36.66,12) and (38,13.34) .. (38,15) .. controls (38,16.66) and (36.66,18) .. (35,18) .. controls (33.34,18) and (32,16.66) .. (32,15) -- cycle ;
\draw  [draw opacity=0] (10,10) -- (40,10) -- (40,20) -- (10,20) -- cycle ;

\end{tikzpicture}} & Load & Bus & $P$, $Q$, $I$, $\mathring{P}$, $\mathring{Q}$ & - \\
        - & Battery & Bus & $P$, $Q$, $I$, $\mathring{P}$, $\mathring{Q}$, $\mathring{V}$, $\overline{P}$, $\underline{P}$, $\overline{Q}$, $\underline{Q}$, Regulation Mode & - \\
        - & SVC & Bus & $P$, $Q$, $I$, $\mathring{V}$, $\mathring{Q}$, Regulation Mode, etc. & - \\
        - & VSC Stations & Station, Bus & $P$, $Q$, $I$, $\mathring{V}$, $\mathring{Q}$, $\underline{Q}$, $\overline{Q}$, Regulation Mode, etc. & - \\                
        - & HVDC Line & Station$_1$, Station$_2$ & $\mathring{P}$, $\overline{P}$, $R$, Droop, etc. & - \\
        \hline
        \resizebox{0.04\linewidth}{!}{\tikzset{every picture/.style={line width=0.75pt}} 

\begin{tikzpicture}[x=0.75pt,y=0.75pt,yscale=-1,xscale=1]

\draw [line width=1.5]    (55,49) -- (45,45) ;
\draw [line width=1.5]    (35,45) -- (40,45) ;
\draw  [draw opacity=0][fill={rgb, 255:red, 0; green, 0; blue, 0 }  ,fill opacity=1 ] (52,49) .. controls (52,47.34) and (53.34,46) .. (55,46) .. controls (56.66,46) and (58,47.34) .. (58,49) .. controls (58,50.66) and (56.66,52) .. (55,52) .. controls (53.34,52) and (52,50.66) .. (52,49) -- cycle ;
\draw  [draw opacity=0][fill={rgb, 255:red, 0; green, 0; blue, 0 }  ,fill opacity=1 ] (32,45) .. controls (32,43.34) and (33.34,42) .. (35,42) .. controls (36.66,42) and (38,43.34) .. (38,45) .. controls (38,46.66) and (36.66,48) .. (35,48) .. controls (33.34,48) and (32,46.66) .. (32,45) -- cycle ;
\draw  [draw opacity=0][fill={rgb, 255:red, 0; green, 0; blue, 0 }  ,fill opacity=1 ] (52,41) .. controls (52,39.34) and (53.34,38) .. (55,38) .. controls (56.66,38) and (58,39.34) .. (58,41) .. controls (58,42.66) and (56.66,44) .. (55,44) .. controls (53.34,44) and (52,42.66) .. (52,41) -- cycle ;
\draw  [draw opacity=0] (30,40) -- (60,40) -- (60,50) -- (30,50) -- cycle ;
\draw [line width=1.5]    (45,45) -- (55,41) ;
\draw  [fill={rgb, 255:red, 255; green, 255; blue, 255 }  ,fill opacity=1 ][line width=1.5]  (45,40) -- (50,45) -- (45,50) -- (40,45) -- cycle ;

\end{tikzpicture}} & Line & Line, Bus$_1$, Bus$_2$ & $P_1$, $Q_1$, $I_1$, $P_2$, $Q_2$, $I_2$, $R$, $X$, $G$, $B$, $\overline{I}_1$, $\overline{I}_2$, $\mathbb{1}_\text{opt}$ & - \\
        \resizebox{0.04\linewidth}{!}{\tikzset{every picture/.style={line width=0.75pt}} 

\begin{tikzpicture}[x=0.75pt,y=0.75pt,yscale=-1,xscale=1]

\draw [color={rgb, 255:red, 208; green, 2; blue, 27 }  ,draw opacity=1 ][line width=1.5]    (35,25) -- (30,25) ;
\draw [color={rgb, 255:red, 208; green, 2; blue, 27 }  ,draw opacity=1 ][line width=1.5]    (20,30) -- (30,20) ;
\draw  [color={rgb, 255:red, 208; green, 2; blue, 27 }  ,draw opacity=1 ][line width=1.5]  (25,20) -- (30,25) -- (25,30) -- (20,25) -- cycle ;
\draw  [draw opacity=0][fill={rgb, 255:red, 0; green, 0; blue, 0 }  ,fill opacity=1 ] (32,25) .. controls (32,23.34) and (33.34,22) .. (35,22) .. controls (36.66,22) and (38,23.34) .. (38,25) .. controls (38,26.66) and (36.66,28) .. (35,28) .. controls (33.34,28) and (32,26.66) .. (32,25) -- cycle ;
\draw  [draw opacity=0] (10,20) -- (40,20) -- (40,30) -- (10,30) -- cycle ;

\end{tikzpicture}} & Line Controller & Line & - & Connected \\
        \hline
        \resizebox{0.04\linewidth}{!}{\tikzset{every picture/.style={line width=0.75pt}} 

\begin{tikzpicture}[x=0.75pt,y=0.75pt,yscale=-1,xscale=1]

\draw [color={rgb, 255:red, 0; green, 0; blue, 0 }  ,draw opacity=1 ][line width=1.5]    (25,25) -- (33,25) ;
\draw [color={rgb, 255:red, 0; green, 0; blue, 0 }  ,draw opacity=1 ][line width=1.5]    (33,20) -- (33,30) ;
\draw [color={rgb, 255:red, 0; green, 0; blue, 0 }  ,draw opacity=1 ][line width=1.5]    (37,20) -- (37,30) ;
\draw [color={rgb, 255:red, 0; green, 0; blue, 0 }  ,draw opacity=1 ][line width=1.5]    (37,25) -- (45,25) ;
\draw  [draw opacity=0][fill={rgb, 255:red, 0; green, 0; blue, 0 }  ,fill opacity=1 ] (42,25) .. controls (42,23.34) and (43.34,22) .. (45,22) .. controls (46.66,22) and (48,23.34) .. (48,25) .. controls (48,26.66) and (46.66,28) .. (45,28) .. controls (43.34,28) and (42,26.66) .. (42,25) -- cycle ;
\draw  [draw opacity=0][fill={rgb, 255:red, 0; green, 0; blue, 0 }  ,fill opacity=1 ] (22,25) .. controls (22,23.34) and (23.34,22) .. (25,22) .. controls (26.66,22) and (28,23.34) .. (28,25) .. controls (28,26.66) and (26.66,28) .. (25,28) .. controls (23.34,28) and (22,26.66) .. (22,25) -- cycle ;
\draw  [draw opacity=0] (20,20) -- (50,20) -- (50,30) -- (20,30) -- cycle ;

\end{tikzpicture}} & Shunt & Shunt, Bus & $P$, $Q$, $I$, $G$, $B$, etc. & - \\
        \resizebox{0.04\linewidth}{!}{\tikzset{every picture/.style={line width=0.75pt}} 

\begin{tikzpicture}[x=0.75pt,y=0.75pt,yscale=-1,xscale=1]

\draw [color={rgb, 255:red, 208; green, 2; blue, 27 }  ,draw opacity=1 ][line width=1.5]    (23.33,10) -- (23.33,20) ;
\draw [color={rgb, 255:red, 208; green, 2; blue, 27 }  ,draw opacity=1 ][line width=1.5]    (26.67,10) -- (26.67,20) ;
\draw [color={rgb, 255:red, 208; green, 2; blue, 27 }  ,draw opacity=1 ][line width=1.5]    (30,10) -- (20,20) ;

\draw [color={rgb, 255:red, 208; green, 2; blue, 27 }  ,draw opacity=1 ][line width=1.5]    (27,15) -- (35,15) ;
\draw  [draw opacity=0][fill={rgb, 255:red, 0; green, 0; blue, 0 }  ,fill opacity=1 ] (32,15) .. controls (32,13.34) and (33.34,12) .. (35,12) .. controls (36.66,12) and (38,13.34) .. (38,15) .. controls (38,16.66) and (36.66,18) .. (35,18) .. controls (33.34,18) and (32,16.66) .. (32,15) -- cycle ;
\draw  [draw opacity=0] (10,10) -- (40,10) -- (40,20) -- (10,20) -- cycle ;

\end{tikzpicture}} & Shunt Controller & Shunt & - & Switch Status \\
        \hline
        \resizebox{0.04\linewidth}{!}{\tikzset{every picture/.style={line width=0.75pt}} 

\begin{tikzpicture}[x=0.75pt,y=0.75pt,yscale=-1,xscale=1]

\draw  [color={rgb, 255:red, 0; green, 0; blue, 0 }  ,draw opacity=1 ][line width=1.5]  (20,15) .. controls (20.82,13.72) and (21.6,12.5) .. (22.5,12.5) .. controls (23.4,12.5) and (24.18,13.72) .. (25,15) .. controls (25.82,16.28) and (26.6,17.5) .. (27.5,17.5) .. controls (28.4,17.5) and (29.18,16.28) .. (30,15) .. controls (30,15) and (30,15) .. (30,15) ;
\draw  [color={rgb, 255:red, 0; green, 0; blue, 0 }  ,draw opacity=1 ][line width=1.5]  (20,15) .. controls (20,12.24) and (22.24,10) .. (25,10) .. controls (27.76,10) and (30,12.24) .. (30,15) .. controls (30,17.76) and (27.76,20) .. (25,20) .. controls (22.24,20) and (20,17.76) .. (20,15) -- cycle ;

\draw [line width=1.5]    (30,15) -- (35,15) ;
\draw [line width=1.5]    (15,15) -- (20,15) ;
\draw  [draw opacity=0][fill={rgb, 255:red, 0; green, 0; blue, 0 }  ,fill opacity=1 ] (32,15) .. controls (32,13.34) and (33.34,12) .. (35,12) .. controls (36.66,12) and (38,13.34) .. (38,15) .. controls (38,16.66) and (36.66,18) .. (35,18) .. controls (33.34,18) and (32,16.66) .. (32,15) -- cycle ;
\draw  [draw opacity=0][fill={rgb, 255:red, 0; green, 0; blue, 0 }  ,fill opacity=1 ] (12,15) .. controls (12,13.34) and (13.34,12) .. (15,12) .. controls (16.66,12) and (18,13.34) .. (18,15) .. controls (18,16.66) and (16.66,18) .. (15,18) .. controls (13.34,18) and (12,16.66) .. (12,15) -- cycle ;
\draw  [draw opacity=0] (10,10) -- (40,10) -- (40,20) -- (10,20) -- cycle ;

\end{tikzpicture}} & Generator & Gen, Bus & $P$, $Q$, $I$, $\mathring{P}$, $\mathring{Q}$, $\mathring{V}$, $\overline{Q}$, $\underline{Q}$, Regulation Mode, etc. & - \\
        \resizebox{0.04\linewidth}{!}{\tikzset{every picture/.style={line width=0.75pt}} 

\begin{tikzpicture}[x=0.75pt,y=0.75pt,yscale=-1,xscale=1]

\draw [color={rgb, 255:red, 0; green, 0; blue, 0 }  ,draw opacity=1 ][line width=1.5]    (15,15) -- (35,15) ;
\draw  [color={rgb, 255:red, 0; green, 0; blue, 0 }  ,draw opacity=1 ][fill={rgb, 255:red, 255; green, 255; blue, 255 }  ,fill opacity=1 ][line width=1.5]  (20,15) .. controls (20,12.24) and (22.24,10) .. (25,10) .. controls (27.76,10) and (30,12.24) .. (30,15) .. controls (30,17.76) and (27.76,20) .. (25,20) .. controls (22.24,20) and (20,17.76) .. (20,15) -- cycle ;
\draw  [color={rgb, 255:red, 0; green, 0; blue, 0 }  ,draw opacity=1 ][fill={rgb, 255:red, 255; green, 255; blue, 255 }  ,fill opacity=1 ][line width=1.5]  (25,11.37) -- (26.15,13.39) -- (28.39,13.88) -- (26.87,15.62) -- (27.1,17.93) -- (25,17) -- (22.9,17.93) -- (23.13,15.62) -- (21.61,13.88) -- (23.85,13.39) -- cycle ;

\draw  [draw opacity=0][fill={rgb, 255:red, 0; green, 0; blue, 0 }  ,fill opacity=1 ] (32,15) .. controls (32,13.34) and (33.34,12) .. (35,12) .. controls (36.66,12) and (38,13.34) .. (38,15) .. controls (38,16.66) and (36.66,18) .. (35,18) .. controls (33.34,18) and (32,16.66) .. (32,15) -- cycle ;
\draw  [draw opacity=0][fill={rgb, 255:red, 0; green, 0; blue, 0 }  ,fill opacity=1 ] (12,15) .. controls (12,13.34) and (13.34,12) .. (15,12) .. controls (16.66,12) and (18,13.34) .. (18,15) .. controls (18,16.66) and (16.66,18) .. (15,18) .. controls (13.34,18) and (12,16.66) .. (12,15) -- cycle ;
\draw  [draw opacity=0] (10,10) -- (40,10) -- (40,20) -- (10,20) -- cycle ;

\end{tikzpicture}} & SVR Unit & Gen, Zone & Participate & - \\
        \resizebox{0.04\linewidth}{!}{\tikzset{every picture/.style={line width=0.75pt}} 

\begin{tikzpicture}[x=0.75pt,y=0.75pt,yscale=-1,xscale=1]

\draw [color={rgb, 255:red, 0; green, 0; blue, 0 }  ,draw opacity=1 ][line width=1.5]    (155,45) -- (175,45) ;
\draw  [draw opacity=0][fill={rgb, 255:red, 0; green, 0; blue, 0 }  ,fill opacity=1 ] (172,45) .. controls (172,43.34) and (173.34,42) .. (175,42) .. controls (176.66,42) and (178,43.34) .. (178,45) .. controls (178,46.66) and (176.66,48) .. (175,48) .. controls (173.34,48) and (172,46.66) .. (172,45) -- cycle ;
\draw  [draw opacity=0][fill={rgb, 255:red, 0; green, 0; blue, 0 }  ,fill opacity=1 ] (152,45) .. controls (152,43.34) and (153.34,42) .. (155,42) .. controls (156.66,42) and (158,43.34) .. (158,45) .. controls (158,46.66) and (156.66,48) .. (155,48) .. controls (153.34,48) and (152,46.66) .. (152,45) -- cycle ;
\draw  [color={rgb, 255:red, 0; green, 0; blue, 0 }  ,draw opacity=1 ][fill={rgb, 255:red, 255; green, 255; blue, 255 }  ,fill opacity=1 ][line width=1.5]  (165,40) -- (166.47,42.98) -- (169.76,43.45) -- (167.38,45.77) -- (167.94,49.05) -- (165,47.5) -- (162.06,49.05) -- (162.62,45.77) -- (160.24,43.45) -- (163.53,42.98) -- cycle ;
\draw  [draw opacity=0] (150,40) -- (180,40) -- (180,50) -- (150,50) -- cycle ;

\end{tikzpicture}} & SVR Zone & Zone, Regulated Bus & $V$, $\vartheta$, $V_\text{nom}$, $\mathring{V}$ & - \\
        \resizebox{0.04\linewidth}{!}{\tikzset{every picture/.style={line width=0.75pt}} 

\begin{tikzpicture}[x=0.75pt,y=0.75pt,yscale=-1,xscale=1]

\draw [color={rgb, 255:red, 208; green, 2; blue, 27 }  ,draw opacity=1 ][line width=1.5]    (25,15) -- (35,15) ;
\draw  [draw opacity=0][fill={rgb, 255:red, 0; green, 0; blue, 0 }  ,fill opacity=1 ] (32,15) .. controls (32,13.34) and (33.34,12) .. (35,12) .. controls (36.66,12) and (38,13.34) .. (38,15) .. controls (38,16.66) and (36.66,18) .. (35,18) .. controls (33.34,18) and (32,16.66) .. (32,15) -- cycle ;
\draw  [color={rgb, 255:red, 208; green, 2; blue, 27 }  ,draw opacity=1 ][fill={rgb, 255:red, 255; green, 255; blue, 255 }  ,fill opacity=1 ][line width=1.5]  (25,10) -- (26.47,12.98) -- (29.76,13.45) -- (27.38,15.77) -- (27.94,19.05) -- (25,17.5) -- (22.06,19.05) -- (22.62,15.77) -- (20.24,13.45) -- (23.53,12.98) -- cycle ;
\draw [color={rgb, 255:red, 208; green, 2; blue, 27 }  ,draw opacity=1 ][line width=1.5]    (20,20) -- (30,10) ;
\draw  [draw opacity=0] (10,10) -- (40,10) -- (40,20) -- (10,20) -- cycle ;

\end{tikzpicture}} & SVR Controller & Zone & - & $\Delta \mathring{V}$ \\
        \hline
        \resizebox{0.04\linewidth}{!}{\tikzset{every picture/.style={line width=0.75pt}} 

\begin{tikzpicture}[x=0.75pt,y=0.75pt,yscale=-1,xscale=1]

\draw [line width=1.5]    (115,65) -- (125,61) ;
\draw [line width=1.5]    (116.67,65.67) -- (125,69) ;
\draw [color={rgb, 255:red, 0; green, 0; blue, 0 }  ,draw opacity=1 ][line width=1.5]    (105,65) -- (115,65) ;
\draw  [fill={rgb, 255:red, 255; green, 255; blue, 255 }  ,fill opacity=1 ][line width=1.5]  (113.33,62) .. controls (113.94,62) and (114.51,62.16) .. (115,62.45) .. controls (115.49,62.16) and (116.06,62) .. (116.67,62) .. controls (118.51,62) and (120,63.49) .. (120,65.33) .. controls (120,67.17) and (118.51,68.67) .. (116.67,68.67) .. controls (116.06,68.67) and (115.49,68.5) .. (115,68.22) .. controls (114.51,68.5) and (113.94,68.67) .. (113.33,68.67) .. controls (111.49,68.67) and (110,67.17) .. (110,65.33) .. controls (110,63.49) and (111.49,62) .. (113.33,62) -- cycle ;
\draw  [fill={rgb, 255:red, 255; green, 255; blue, 255 }  ,fill opacity=1 ][line width=1.5]  (110,65.33) .. controls (110,63.49) and (111.49,62) .. (113.33,62) .. controls (115.17,62) and (116.67,63.49) .. (116.67,65.33) .. controls (116.67,67.17) and (115.17,68.67) .. (113.33,68.67) .. controls (111.49,68.67) and (110,67.17) .. (110,65.33) -- cycle ;

\draw  [draw opacity=0][fill={rgb, 255:red, 0; green, 0; blue, 0 }  ,fill opacity=1 ] (122,69) .. controls (122,67.34) and (123.34,66) .. (125,66) .. controls (126.66,66) and (128,67.34) .. (128,69) .. controls (128,70.66) and (126.66,72) .. (125,72) .. controls (123.34,72) and (122,70.66) .. (122,69) -- cycle ;
\draw  [draw opacity=0][fill={rgb, 255:red, 0; green, 0; blue, 0 }  ,fill opacity=1 ] (102,65) .. controls (102,63.34) and (103.34,62) .. (105,62) .. controls (106.66,62) and (108,63.34) .. (108,65) .. controls (108,66.66) and (106.66,68) .. (105,68) .. controls (103.34,68) and (102,66.66) .. (102,65) -- cycle ;
\draw  [draw opacity=0][fill={rgb, 255:red, 0; green, 0; blue, 0 }  ,fill opacity=1 ] (122,61) .. controls (122,59.34) and (123.34,58) .. (125,58) .. controls (126.66,58) and (128,59.34) .. (128,61) .. controls (128,62.66) and (126.66,64) .. (125,64) .. controls (123.34,64) and (122,62.66) .. (122,61) -- cycle ;
\draw  [draw opacity=0] (100,60) -- (130,60) -- (130,70) -- (100,70) -- cycle ;

\end{tikzpicture}} & TWT & TWT, Bus$_1$, Bus$_2$ & $P_1$, $Q_1$, $I_1$, $P_2$, $Q_2$, $I_2$, $R$, $X$, $G$, $B$, $\rho$, $\alpha$, $\overline{I}_1$, $\overline{I}_2$, $\mathbb{1}_\text{opt}$ & - \\               
        \resizebox{0.04\linewidth}{!}{\tikzset{every picture/.style={line width=0.75pt}} 

\begin{tikzpicture}[x=0.75pt,y=0.75pt,yscale=-1,xscale=1]

\draw [color={rgb, 255:red, 0; green, 0; blue, 0 }  ,draw opacity=1 ][line width=1.5]    (15,15) -- (35,15) ;
\draw  [draw opacity=0][fill={rgb, 255:red, 0; green, 0; blue, 0 }  ,fill opacity=1 ] (32,15) .. controls (32,13.34) and (33.34,12) .. (35,12) .. controls (36.66,12) and (38,13.34) .. (38,15) .. controls (38,16.66) and (36.66,18) .. (35,18) .. controls (33.34,18) and (32,16.66) .. (32,15) -- cycle ;
\draw  [draw opacity=0][fill={rgb, 255:red, 0; green, 0; blue, 0 }  ,fill opacity=1 ] (12,15) .. controls (12,13.34) and (13.34,12) .. (15,12) .. controls (16.66,12) and (18,13.34) .. (18,15) .. controls (18,16.66) and (16.66,18) .. (15,18) .. controls (13.34,18) and (12,16.66) .. (12,15) -- cycle ;
\draw  [draw opacity=0] (10,10) -- (40,10) -- (40,20) -- (10,20) -- cycle ;
\draw  [color={rgb, 255:red, 0; green, 0; blue, 0 }  ,draw opacity=1 ][fill={rgb, 255:red, 255; green, 255; blue, 255 }  ,fill opacity=1 ][line width=1.5]  (20,12) -- (26.67,12) -- (26.67,18.67) -- (20,18.67) -- cycle ;
\draw  [color={rgb, 255:red, 0; green, 0; blue, 0 }  ,draw opacity=1 ][fill={rgb, 255:red, 255; green, 255; blue, 255 }  ,fill opacity=1 ][line width=1.5]  (26.67,12) -- (30,12) -- (30,18.67) -- (26.67,18.67) -- cycle ;

\end{tikzpicture}} & RTC & TWT, Regulated Bus & - & - \\
        \resizebox{0.04\linewidth}{!}{\tikzset{every picture/.style={line width=0.75pt}} 

\begin{tikzpicture}[x=0.75pt,y=0.75pt,yscale=-1,xscale=1]

\draw [color={rgb, 255:red, 208; green, 2; blue, 27 }  ,draw opacity=1 ][line width=1.5]    (25,15) -- (35,15) ;
\draw  [draw opacity=0][fill={rgb, 255:red, 0; green, 0; blue, 0 }  ,fill opacity=1 ] (32,15) .. controls (32,13.34) and (33.34,12) .. (35,12) .. controls (36.66,12) and (38,13.34) .. (38,15) .. controls (38,16.66) and (36.66,18) .. (35,18) .. controls (33.34,18) and (32,16.66) .. (32,15) -- cycle ;
\draw  [draw opacity=0] (10,10) -- (40,10) -- (40,20) -- (10,20) -- cycle ;
\draw  [color={rgb, 255:red, 208; green, 2; blue, 27 }  ,draw opacity=1 ][fill={rgb, 255:red, 255; green, 255; blue, 255 }  ,fill opacity=1 ][line width=1.5]  (20,12) -- (26.67,12) -- (26.67,18.67) -- (20,18.67) -- cycle ;
\draw  [color={rgb, 255:red, 208; green, 2; blue, 27 }  ,draw opacity=1 ][fill={rgb, 255:red, 255; green, 255; blue, 255 }  ,fill opacity=1 ][line width=1.5]  (26.67,12) -- (30,12) -- (30,18.67) -- (26.67,18.67) -- cycle ;

\draw [color={rgb, 255:red, 208; green, 2; blue, 27 }  ,draw opacity=1 ][line width=1.5]    (30,10) -- (20,20) ;

\end{tikzpicture}} & RTC Controller & TWT & $\mathring{V}$, $V_{nom}$ & $\mathbb{1}_{0\%}$, $\mathbb{1}_{2\%}$, $\mathbb{1}_{5\%}$, $\mathbb{1}_{7\%}$ \\
        \hline
    \end{tabular}
    \caption{List of hyper-edge classes that compose an operating condition $x$.
        Hyper-edges can have one, two or three ports.
        When \textit{etc.} is displayed, context features lists are non exhaustive for the sake of readability.
        They are however mostly compliant with the features used in the PowSyBl framework \cite{powsybl}, at the exception of the $\mathbb{1}_\text{opt}$ feature which identifies whether an object is to be included in the objective function (see eqn. \eqref{eq:objective_function}-\eqref{eq:joule}).
        $P$, $Q$, $I$, $V$ and $\vartheta$ respectively denote active power, reactive power, current, voltage magnitude and phase angle after a first AC power flow simulation.
        $\mathring{\cdot}$ denotes a target setpoint, while $\underline{\cdot}$ and $\overline{\cdot}$ denote low and high limits.
        $V_{nom}$ denotes a nominal voltage.
        $R$, $X$, $G$ and $B$ respectively denote resistance, reactance, conductance and susceptance.
        In the context of this table, $\rho$ and $\alpha$ denote transformation ratio and phase shift.
        Indices $1$ and $2$ denote origin and extremity ends when applicable.
        }
    \label{tab:classes}
\end{table*}

    \subsection{Related Literature}

    By means of several more or less reasonable assumptions, the TVC problem may be framed as an \emph{AC Optimal Power Flow} (AC-OPF) \cite{Castillo2016} problem, taking the form of a \emph{Mixed-Integer Non Linear Program} (MINLP) \cite{Fukuyama}.
    Unfortunately, the significant size of the system (27 continuous and 882 discrete decision variables in our experiments) make state-of-the-art approaches incompatible with real-time and full-scale decision-making.
    Thus, the \emph{Power Systems} community has recently started exploring the use of methods from the \emph{Deep Learning} (DL) literature \cite{goodfellow2016deep}, which allow for very short computational times, at the cost of a potentially expensive training phase.
    In the case of TVC, even if recommended actions come without any sort of optimality guarantee, they can always be checked for security by a power system simulator before applying them in real operation.
    
    \emph{Deep Neural Networks} (DNNs) are a class of highly expressive parametric mappings capable of approximating any function \cite{hornik1991approximation}.
    Directly training a DNN to imitate operators is made impractical by the wide variety of observed behaviors.
    Similarly, the imitation of a MINLP solver requires a large amount of labeled data which cannot be generated on real-life systems in a reasonable amount of time.
    A third idea is to train a DNN in a self-supervised fashion, letting it discover by trial and error interactions with a power system digital simulator which are the most relevant actions in a given context.
    Some previous work \cite{9320077, hagmar2022, ZHEN202243} frame the AC-OPF as a closed-loop \emph{Reinforcement Learning} (RL) \cite{sutton} problem, while others \cite{pan2022} acknowledge its open-loop and deterministic nature, falling in line with the \emph{Amortized Optimization} (AO) literature \cite{amortizedoptimization}.
    
    Real-life power grid operating conditions are \emph{Hyper Heterogeneous Multi Graphs} (H2MGs) \cite{donon2022}, whose topological structures largely vary from one operating condition to the other \cite{RTE7k}, whether it be because of maintenance, bus-splitting or objects renaming.
    They cannot be properly handled by most DNN architectures, which assume a constant number and ordering of objects that compose the grid.
    Meanwhile, \emph{Graph Neural Networks} (GNNs) \cite{gnn_original, thomaskipfgnn} are a class of DNNs especially designed to handle graph data and are robust to object addition, removal and reordering, making them natural candidates for power systems applications \cite{liao2021reviewgraphneuralnetworks, lopezcardona2025, owerko2022, li2022, diehl2019}.
    The present work extends our previously-introduced GNN architecture called \emph{Hyper Heterogeneous Multi Graph Neural Ordinary Differential Equation} (H2MGNODE) \cite{donon2024}, so as to natively process real-life and full-scale data from the French HV-EHV system without any substantial preprocessing (at the exception of a rescaling of input features).    

    \subsection{Contributions}

    The present work specifically considers the TVC problem in the context of post-processing the existing forecasting pipeline, a critical tool to anticipate issues and decide on preventive actions.
    Although accurate in terms of active power injections, the forecasting pipeline does very little to predict realistic voltages, as it would require a somewhat faithful proxy of operators' TVC decisions.
    As a consequence, operation planners are left unaware of incoming voltage issues that they could have anticipated.
    We aim at training a GNN to find the best actions to alleviate most voltage issues, so that operators can focus on the ones that require their careful attention.

    Major contributions of the present study include:
    \begin{itemize}[noitemsep,topsep=0pt]
        \item A faithful and generic, H2MG-based, representation of the cyber-physical model of real-life power grids;
        \item A comprehensive, GNN-based, AO methodology for the self-supervised training of a TVC policy by interacting with an industrial 
        power system simulator.
        \item A large-scale experimental validation of the approach via the training/validation/testing of a model on observational data (about 170,000 operating states) of the full HV-EHV French grid (about 7,000 buses), gathered over one year of day-ahead forecasting activities.
    \end{itemize}

    \subsection{Paper Organization}
    The rest of this paper is organized as follows.
    Section \ref{sec:methodology} defines the practical optimization problem and transforms it into an AO problem, Section \ref{sec:case_study} details the experimental protocol and results, and Section \ref{sec:conclusion} summarizes the main findings and outlines the next steps required to reach the deployment of our proposed methodology.

\section{Methodology}
\label{sec:methodology}

Let us detail how we transform the initial black-box optimization problem into an AO problem.

    \subsection{Initial Optimization Problem}

    This work considers the problem of improving voltages using a series of levers of different natures, spread across a real-life cyber-physical system.
    To ensure a seamless and faithful representation of the real-life data at hand, we choose to frame operating conditions as \emph{Hyper Heterogeneous Multi Graphs} (H2MGs) \cite{donon2022}.

\begin{figure*}[b!]
            \centering
            \begin{subfigure}[b]{0.48\linewidth}
                \centering
                \resizebox{0.85\linewidth}{!}{\tikzset{every picture/.style={line width=0.75pt}} 

\begin{tikzpicture}[x=0.75pt,y=0.75pt,yscale=-1,xscale=1]

\draw [color={rgb, 255:red, 208; green, 2; blue, 27 }  ,draw opacity=1 ][line width=1.5]    (105,30) -- (105,35) -- (75,35) -- (75,30) ;
\draw  [color={rgb, 255:red, 0; green, 0; blue, 0 }  ,draw opacity=1 ][line width=1.5]  (83,74.72) .. controls (84.09,73) and (85.13,71.36) .. (86.33,71.36) .. controls (87.54,71.36) and (88.58,73) .. (89.67,74.72) .. controls (90.75,76.44) and (91.79,78.08) .. (93,78.08) .. controls (94.21,78.08) and (95.25,76.44) .. (96.33,74.72) .. controls (96.33,74.72) and (96.33,74.72) .. (96.33,74.72) ;
\draw  [color={rgb, 255:red, 0; green, 0; blue, 0 }  ,draw opacity=1 ][line width=1.5]  (83,74.72) .. controls (83,71.01) and (85.98,68) .. (89.67,68) .. controls (93.35,68) and (96.33,71.01) .. (96.33,74.72) .. controls (96.33,78.43) and (93.35,81.44) .. (89.67,81.44) .. controls (85.98,81.44) and (83,78.43) .. (83,74.72) -- cycle ;

\draw [color={rgb, 255:red, 0; green, 0; blue, 0 }  ,draw opacity=1 ][line width=1.5]    (89.67,60.94) -- (89.67,68) ;
\draw [color={rgb, 255:red, 0; green, 0; blue, 0 }  ,draw opacity=1 ][line width=3]    (100,30) -- (120,30) ;
\draw [color={rgb, 255:red, 0; green, 0; blue, 0 }  ,draw opacity=1 ][line width=1.5]    (115,30) -- (115,39) -- (95,55) -- (95,60) ;
\draw [color={rgb, 255:red, 0; green, 0; blue, 0 }  ,draw opacity=1 ][line width=1.5]    (65,30) -- (65,40) -- (85,55) -- (85,60) ;
\draw [color={rgb, 255:red, 0; green, 0; blue, 0 }  ,draw opacity=1 ][line width=3]    (60,30) -- (80,30) ;
\draw [color={rgb, 255:red, 0; green, 0; blue, 0 }  ,draw opacity=1 ][line width=3]    (80,60) -- (100,60) ;
\draw [color={rgb, 255:red, 0; green, 0; blue, 0 }  ,draw opacity=1 ][line width=1.5]    (110,30) -- (110,13) ;
\draw [shift={(110,10)}, rotate = 90] [color={rgb, 255:red, 0; green, 0; blue, 0 }  ,draw opacity=1 ][line width=1.5]    (14.21,-4.28) .. controls (9.04,-1.82) and (4.3,-0.39) .. (0,0) .. controls (4.3,0.39) and (9.04,1.82) .. (14.21,4.28)   ;
\draw  [line width=1.5]  (200,30) -- (210,30) -- (210,40) -- (200,40) -- cycle ;
\draw [line width=1.5]    (200,30) -- (210,40) ;
\draw [line width=1.5]    (200,40) -- (210,30) ;

\draw  [color={rgb, 255:red, 0; green, 0; blue, 0 }  ,draw opacity=1 ][line width=1.5]  (235,75) .. controls (235.82,73.72) and (236.6,72.5) .. (237.5,72.5) .. controls (238.4,72.5) and (239.18,73.72) .. (240,75) .. controls (240.82,76.28) and (241.6,77.5) .. (242.5,77.5) .. controls (243.4,77.5) and (244.18,76.28) .. (245,75) .. controls (245,75) and (245,75) .. (245,75) ;
\draw  [color={rgb, 255:red, 0; green, 0; blue, 0 }  ,draw opacity=1 ][line width=1.5]  (235,75) .. controls (235,72.24) and (237.24,70) .. (240,70) .. controls (242.76,70) and (245,72.24) .. (245,75) .. controls (245,77.76) and (242.76,80) .. (240,80) .. controls (237.24,80) and (235,77.76) .. (235,75) -- cycle ;

\draw [line width=1.5]    (244,71) -- (250,65) ;
\draw  [draw opacity=0][fill={rgb, 255:red, 0; green, 0; blue, 0 }  ,fill opacity=1 ] (247,65) .. controls (247,63.34) and (248.34,62) .. (250,62) .. controls (251.66,62) and (253,63.34) .. (253,65) .. controls (253,66.66) and (251.66,68) .. (250,68) .. controls (248.34,68) and (247,66.66) .. (247,65) -- cycle ;
\draw  [draw opacity=0][fill={rgb, 255:red, 0; green, 0; blue, 0 }  ,fill opacity=1 ] (224,75) .. controls (224,73.34) and (225.34,72) .. (227,72) .. controls (228.66,72) and (230,73.34) .. (230,75) .. controls (230,76.66) and (228.66,78) .. (227,78) .. controls (225.34,78) and (224,76.66) .. (224,75) -- cycle ;
\draw [line width=1.5]    (268,53) -- (273,58) ;
\draw  [line width=1.5]  (265,45) -- (270,50) -- (265,55) -- (260,50) -- cycle ;
\draw [line width=1.5]    (280,35) -- (268,47) ;
\draw [line width=1.5]    (262,53) -- (250,65) ;
\draw  [draw opacity=0][fill={rgb, 255:red, 0; green, 0; blue, 0 }  ,fill opacity=1 ] (217,35) .. controls (217,33.34) and (218.34,32) .. (220,32) .. controls (221.66,32) and (223,33.34) .. (223,35) .. controls (223,36.66) and (221.66,38) .. (220,38) .. controls (218.34,38) and (217,36.66) .. (217,35) -- cycle ;
\draw  [draw opacity=0][fill={rgb, 255:red, 0; green, 0; blue, 0 }  ,fill opacity=1 ] (270,58) .. controls (270,56.34) and (271.34,55) .. (273,55) .. controls (274.66,55) and (276,56.34) .. (276,58) .. controls (276,59.66) and (274.66,61) .. (273,61) .. controls (271.34,61) and (270,59.66) .. (270,58) -- cycle ;
\draw [line width=1.5]    (227,75) -- (235,75) ;
\draw [color={rgb, 255:red, 0; green, 0; blue, 0 }  ,draw opacity=1 ][line width=3]    (255,75) -- (265,75) ;
\draw [line width=1.5]    (250,65) -- (260,75) ;
\draw [color={rgb, 255:red, 0; green, 0; blue, 0 }  ,draw opacity=1 ][line width=3]    (215,45) -- (225,45) ;
\draw [line width=1.5]    (220,35) -- (220,45) ;
\draw [color={rgb, 255:red, 0; green, 0; blue, 0 }  ,draw opacity=1 ][line width=3]    (275,45) -- (285,45) ;
\draw [line width=1.5]    (280,35) -- (280,45) ;
\draw  [line width=1.5]  (235,45) -- (240,50) -- (235,55) -- (230,50) -- cycle ;
\draw [line width=1.5]    (220,35) -- (232,47) ;
\draw  [draw opacity=0][fill={rgb, 255:red, 0; green, 0; blue, 0 }  ,fill opacity=1 ] (277,35) .. controls (277,33.34) and (278.34,32) .. (280,32) .. controls (281.66,32) and (283,33.34) .. (283,35) .. controls (283,36.66) and (281.66,38) .. (280,38) .. controls (278.34,38) and (277,36.66) .. (277,35) -- cycle ;
\draw [line width=1.5]    (238,53) -- (250,65) ;
\draw [line width=1.5]    (232,53) -- (227,58) ;
\draw  [draw opacity=0][fill={rgb, 255:red, 0; green, 0; blue, 0 }  ,fill opacity=1 ] (224,58) .. controls (224,56.34) and (225.34,55) .. (227,55) .. controls (228.66,55) and (230,56.34) .. (230,58) .. controls (230,59.66) and (228.66,61) .. (227,61) .. controls (225.34,61) and (224,59.66) .. (224,58) -- cycle ;
\draw  [line width=1.5]  (250,30) -- (255,35) -- (250,40) -- (245,35) -- cycle ;
\draw [line width=1.5]    (220,35) -- (245,35) ;
\draw [line width=1.5]    (255,35) -- (280,35) ;
\draw [line width=1.5]    (250,25) -- (250,30) ;
\draw [line width=1.5]    (210,35) -- (220,35) ;
\draw [line width=1.5]    (280,35) -- (290,35) ;
\draw  [line width=1.5]  (290,30) -- (300,30) -- (300,40) -- (290,40) -- cycle ;
\draw [line width=1.5]    (290,30) -- (300,40) ;
\draw [line width=1.5]    (290,40) -- (300,30) ;

\draw [color={rgb, 255:red, 208; green, 2; blue, 27 }  ,draw opacity=1 ][line width=1.5]    (250,25) -- (250,20) ;
\draw  [draw opacity=0][fill={rgb, 255:red, 0; green, 0; blue, 0 }  ,fill opacity=1 ] (247,25) .. controls (247,23.34) and (248.34,22) .. (250,22) .. controls (251.66,22) and (253,23.34) .. (253,25) .. controls (253,26.66) and (251.66,28) .. (250,28) .. controls (248.34,28) and (247,26.66) .. (247,25) -- cycle ;
\draw [color={rgb, 255:red, 208; green, 2; blue, 27 }  ,draw opacity=1 ][line width=1.5]    (245,20) -- (255,10) ;
\draw  [color={rgb, 255:red, 208; green, 2; blue, 27 }  ,draw opacity=1 ][line width=1.5]  (250,10) -- (255,15) -- (250,20) -- (245,15) -- cycle ;
\draw [color={rgb, 255:red, 0; green, 0; blue, 0 }  ,draw opacity=1 ][line width=1.5]    (70,30) -- (70,13) ;
\draw [shift={(70,10)}, rotate = 90] [color={rgb, 255:red, 0; green, 0; blue, 0 }  ,draw opacity=1 ][line width=1.5]    (14.21,-4.28) .. controls (9.04,-1.82) and (4.3,-0.39) .. (0,0) .. controls (4.3,0.39) and (9.04,1.82) .. (14.21,4.28)   ;
\draw  [draw opacity=0] (20,10) -- (320,10) -- (320,80) -- (20,80) -- cycle ;

\end{tikzpicture}}
                \caption{
                    \textit{Line Controller} example.
                    In the H2MG representation (right), \textit{Lines} (\resizebox{0.08\linewidth}{!}{\tikzset{every picture/.style={line width=0.75pt}} 

\begin{tikzpicture}[x=0.75pt,y=0.75pt,yscale=-1,xscale=1]

\draw [line width=1.5]    (55,49) -- (45,45) ;
\draw [line width=1.5]    (35,45) -- (40,45) ;
\draw  [draw opacity=0][fill={rgb, 255:red, 0; green, 0; blue, 0 }  ,fill opacity=1 ] (52,49) .. controls (52,47.34) and (53.34,46) .. (55,46) .. controls (56.66,46) and (58,47.34) .. (58,49) .. controls (58,50.66) and (56.66,52) .. (55,52) .. controls (53.34,52) and (52,50.66) .. (52,49) -- cycle ;
\draw  [draw opacity=0][fill={rgb, 255:red, 0; green, 0; blue, 0 }  ,fill opacity=1 ] (32,45) .. controls (32,43.34) and (33.34,42) .. (35,42) .. controls (36.66,42) and (38,43.34) .. (38,45) .. controls (38,46.66) and (36.66,48) .. (35,48) .. controls (33.34,48) and (32,46.66) .. (32,45) -- cycle ;
\draw  [draw opacity=0][fill={rgb, 255:red, 0; green, 0; blue, 0 }  ,fill opacity=1 ] (52,41) .. controls (52,39.34) and (53.34,38) .. (55,38) .. controls (56.66,38) and (58,39.34) .. (58,41) .. controls (58,42.66) and (56.66,44) .. (55,44) .. controls (53.34,44) and (52,42.66) .. (52,41) -- cycle ;
\draw  [draw opacity=0] (30,40) -- (60,40) -- (60,50) -- (30,50) -- cycle ;
\draw [line width=1.5]    (45,45) -- (55,41) ;
\draw  [fill={rgb, 255:red, 255; green, 255; blue, 255 }  ,fill opacity=1 ][line width=1.5]  (45,40) -- (50,45) -- (45,50) -- (40,45) -- cycle ;

\end{tikzpicture}}) are 3$^\text{rd}$ order hyper-edges.
                    Controllable \textit{Lines} -- such as the red one in the SLD representation (left) -- are connected via their \textit{Line} port to \textit{Line Controllers} (\resizebox{0.08\linewidth}{!}{\tikzset{every picture/.style={line width=0.75pt}} 

\begin{tikzpicture}[x=0.75pt,y=0.75pt,yscale=-1,xscale=1]

\draw [color={rgb, 255:red, 208; green, 2; blue, 27 }  ,draw opacity=1 ][line width=1.5]    (35,25) -- (30,25) ;
\draw [color={rgb, 255:red, 208; green, 2; blue, 27 }  ,draw opacity=1 ][line width=1.5]    (20,30) -- (30,20) ;
\draw  [color={rgb, 255:red, 208; green, 2; blue, 27 }  ,draw opacity=1 ][line width=1.5]  (25,20) -- (30,25) -- (25,30) -- (20,25) -- cycle ;
\draw  [draw opacity=0][fill={rgb, 255:red, 0; green, 0; blue, 0 }  ,fill opacity=1 ] (32,25) .. controls (32,23.34) and (33.34,22) .. (35,22) .. controls (36.66,22) and (38,23.34) .. (38,25) .. controls (38,26.66) and (36.66,28) .. (35,28) .. controls (33.34,28) and (32,26.66) .. (32,25) -- cycle ;
\draw  [draw opacity=0] (10,20) -- (40,20) -- (40,30) -- (10,30) -- cycle ;

\end{tikzpicture}}), which are 1$^\text{st}$ order hyper-edges that bear the decision variable \textit{Connected}.
                }
                \label{fig:line_all}
            \end{subfigure}
            \hfill
            \begin{subfigure}[b]{0.48\linewidth}
                \centering
                \resizebox{0.85\linewidth}{!}{\input{shunt_all.tikz}}
                \caption{
                    \textit{Shunt Controller} example.
                    In the H2MG representation (right), \textit{Shunts} (\resizebox{0.08\linewidth}{!}{\tikzset{every picture/.style={line width=0.75pt}} 

\begin{tikzpicture}[x=0.75pt,y=0.75pt,yscale=-1,xscale=1]

\draw [color={rgb, 255:red, 0; green, 0; blue, 0 }  ,draw opacity=1 ][line width=1.5]    (25,25) -- (33,25) ;
\draw [color={rgb, 255:red, 0; green, 0; blue, 0 }  ,draw opacity=1 ][line width=1.5]    (33,20) -- (33,30) ;
\draw [color={rgb, 255:red, 0; green, 0; blue, 0 }  ,draw opacity=1 ][line width=1.5]    (37,20) -- (37,30) ;
\draw [color={rgb, 255:red, 0; green, 0; blue, 0 }  ,draw opacity=1 ][line width=1.5]    (37,25) -- (45,25) ;
\draw  [draw opacity=0][fill={rgb, 255:red, 0; green, 0; blue, 0 }  ,fill opacity=1 ] (42,25) .. controls (42,23.34) and (43.34,22) .. (45,22) .. controls (46.66,22) and (48,23.34) .. (48,25) .. controls (48,26.66) and (46.66,28) .. (45,28) .. controls (43.34,28) and (42,26.66) .. (42,25) -- cycle ;
\draw  [draw opacity=0][fill={rgb, 255:red, 0; green, 0; blue, 0 }  ,fill opacity=1 ] (22,25) .. controls (22,23.34) and (23.34,22) .. (25,22) .. controls (26.66,22) and (28,23.34) .. (28,25) .. controls (28,26.66) and (26.66,28) .. (25,28) .. controls (23.34,28) and (22,26.66) .. (22,25) -- cycle ;
\draw  [draw opacity=0] (20,20) -- (50,20) -- (50,30) -- (20,30) -- cycle ;

\end{tikzpicture}}) are 2$^\text{nd}$ order hyper-edges.
                    Controllable \textit{Shunts} -- such as the red one in the SLD representation (left) -- are connected via their \textit{Shunt} port to \textit{Shunt Controllers} (\resizebox{0.08\linewidth}{!}{\tikzset{every picture/.style={line width=0.75pt}} 

\begin{tikzpicture}[x=0.75pt,y=0.75pt,yscale=-1,xscale=1]

\draw [color={rgb, 255:red, 208; green, 2; blue, 27 }  ,draw opacity=1 ][line width=1.5]    (23.33,10) -- (23.33,20) ;
\draw [color={rgb, 255:red, 208; green, 2; blue, 27 }  ,draw opacity=1 ][line width=1.5]    (26.67,10) -- (26.67,20) ;
\draw [color={rgb, 255:red, 208; green, 2; blue, 27 }  ,draw opacity=1 ][line width=1.5]    (30,10) -- (20,20) ;

\draw [color={rgb, 255:red, 208; green, 2; blue, 27 }  ,draw opacity=1 ][line width=1.5]    (27,15) -- (35,15) ;
\draw  [draw opacity=0][fill={rgb, 255:red, 0; green, 0; blue, 0 }  ,fill opacity=1 ] (32,15) .. controls (32,13.34) and (33.34,12) .. (35,12) .. controls (36.66,12) and (38,13.34) .. (38,15) .. controls (38,16.66) and (36.66,18) .. (35,18) .. controls (33.34,18) and (32,16.66) .. (32,15) -- cycle ;
\draw  [draw opacity=0] (10,10) -- (40,10) -- (40,20) -- (10,20) -- cycle ;

\end{tikzpicture}}), which are 1$^\text{st}$ order hyper-edges that bear the decision variable \textit{Switch Status}.
                }
                \label{fig:shunt_all}
            \end{subfigure}
            \vfill
            \begin{subfigure}[b]{0.48\linewidth}
                \centering
                \resizebox{0.85\linewidth}{!}{\input{rst_all.tikz}}
                \caption{
                    \textit{SVR Controller} example.
                    SVR is a closed-loop regulation that modifies the reactive power of a generator cluster to regulate the voltage of a regulated bus.
                    In the H2MG representation (right), \textit{Generators} (\resizebox{0.08\linewidth}{!}{\tikzset{every picture/.style={line width=0.75pt}} 

\begin{tikzpicture}[x=0.75pt,y=0.75pt,yscale=-1,xscale=1]

\draw  [color={rgb, 255:red, 0; green, 0; blue, 0 }  ,draw opacity=1 ][line width=1.5]  (20,15) .. controls (20.82,13.72) and (21.6,12.5) .. (22.5,12.5) .. controls (23.4,12.5) and (24.18,13.72) .. (25,15) .. controls (25.82,16.28) and (26.6,17.5) .. (27.5,17.5) .. controls (28.4,17.5) and (29.18,16.28) .. (30,15) .. controls (30,15) and (30,15) .. (30,15) ;
\draw  [color={rgb, 255:red, 0; green, 0; blue, 0 }  ,draw opacity=1 ][line width=1.5]  (20,15) .. controls (20,12.24) and (22.24,10) .. (25,10) .. controls (27.76,10) and (30,12.24) .. (30,15) .. controls (30,17.76) and (27.76,20) .. (25,20) .. controls (22.24,20) and (20,17.76) .. (20,15) -- cycle ;

\draw [line width=1.5]    (30,15) -- (35,15) ;
\draw [line width=1.5]    (15,15) -- (20,15) ;
\draw  [draw opacity=0][fill={rgb, 255:red, 0; green, 0; blue, 0 }  ,fill opacity=1 ] (32,15) .. controls (32,13.34) and (33.34,12) .. (35,12) .. controls (36.66,12) and (38,13.34) .. (38,15) .. controls (38,16.66) and (36.66,18) .. (35,18) .. controls (33.34,18) and (32,16.66) .. (32,15) -- cycle ;
\draw  [draw opacity=0][fill={rgb, 255:red, 0; green, 0; blue, 0 }  ,fill opacity=1 ] (12,15) .. controls (12,13.34) and (13.34,12) .. (15,12) .. controls (16.66,12) and (18,13.34) .. (18,15) .. controls (18,16.66) and (16.66,18) .. (15,18) .. controls (13.34,18) and (12,16.66) .. (12,15) -- cycle ;
\draw  [draw opacity=0] (10,10) -- (40,10) -- (40,20) -- (10,20) -- cycle ;

\end{tikzpicture}}) are 2$^\text{nd}$ order hyper-edges.
                    \textit{Generators} taking part in the regulation -- such as the red ones in the SLD representation (left) -- are connected via their \textit{Gen} port to \textit{SVR Units} (\resizebox{0.08\linewidth}{!}{\tikzset{every picture/.style={line width=0.75pt}} 

\begin{tikzpicture}[x=0.75pt,y=0.75pt,yscale=-1,xscale=1]

\draw [color={rgb, 255:red, 0; green, 0; blue, 0 }  ,draw opacity=1 ][line width=1.5]    (15,15) -- (35,15) ;
\draw  [color={rgb, 255:red, 0; green, 0; blue, 0 }  ,draw opacity=1 ][fill={rgb, 255:red, 255; green, 255; blue, 255 }  ,fill opacity=1 ][line width=1.5]  (20,15) .. controls (20,12.24) and (22.24,10) .. (25,10) .. controls (27.76,10) and (30,12.24) .. (30,15) .. controls (30,17.76) and (27.76,20) .. (25,20) .. controls (22.24,20) and (20,17.76) .. (20,15) -- cycle ;
\draw  [color={rgb, 255:red, 0; green, 0; blue, 0 }  ,draw opacity=1 ][fill={rgb, 255:red, 255; green, 255; blue, 255 }  ,fill opacity=1 ][line width=1.5]  (25,11.37) -- (26.15,13.39) -- (28.39,13.88) -- (26.87,15.62) -- (27.1,17.93) -- (25,17) -- (22.9,17.93) -- (23.13,15.62) -- (21.61,13.88) -- (23.85,13.39) -- cycle ;

\draw  [draw opacity=0][fill={rgb, 255:red, 0; green, 0; blue, 0 }  ,fill opacity=1 ] (32,15) .. controls (32,13.34) and (33.34,12) .. (35,12) .. controls (36.66,12) and (38,13.34) .. (38,15) .. controls (38,16.66) and (36.66,18) .. (35,18) .. controls (33.34,18) and (32,16.66) .. (32,15) -- cycle ;
\draw  [draw opacity=0][fill={rgb, 255:red, 0; green, 0; blue, 0 }  ,fill opacity=1 ] (12,15) .. controls (12,13.34) and (13.34,12) .. (15,12) .. controls (16.66,12) and (18,13.34) .. (18,15) .. controls (18,16.66) and (16.66,18) .. (15,18) .. controls (13.34,18) and (12,16.66) .. (12,15) -- cycle ;
\draw  [draw opacity=0] (10,10) -- (40,10) -- (40,20) -- (10,20) -- cycle ;

\end{tikzpicture}}) which are 2$^\text{nd}$ order hyper-edges.
                    The regulated \textit{Bus} is connected via its \textit{Bus} port to an \textit{SVR Zone} (\resizebox{0.08\linewidth}{!}{\tikzset{every picture/.style={line width=0.75pt}} 

\begin{tikzpicture}[x=0.75pt,y=0.75pt,yscale=-1,xscale=1]

\draw [color={rgb, 255:red, 0; green, 0; blue, 0 }  ,draw opacity=1 ][line width=1.5]    (155,45) -- (175,45) ;
\draw  [draw opacity=0][fill={rgb, 255:red, 0; green, 0; blue, 0 }  ,fill opacity=1 ] (172,45) .. controls (172,43.34) and (173.34,42) .. (175,42) .. controls (176.66,42) and (178,43.34) .. (178,45) .. controls (178,46.66) and (176.66,48) .. (175,48) .. controls (173.34,48) and (172,46.66) .. (172,45) -- cycle ;
\draw  [draw opacity=0][fill={rgb, 255:red, 0; green, 0; blue, 0 }  ,fill opacity=1 ] (152,45) .. controls (152,43.34) and (153.34,42) .. (155,42) .. controls (156.66,42) and (158,43.34) .. (158,45) .. controls (158,46.66) and (156.66,48) .. (155,48) .. controls (153.34,48) and (152,46.66) .. (152,45) -- cycle ;
\draw  [color={rgb, 255:red, 0; green, 0; blue, 0 }  ,draw opacity=1 ][fill={rgb, 255:red, 255; green, 255; blue, 255 }  ,fill opacity=1 ][line width=1.5]  (165,40) -- (166.47,42.98) -- (169.76,43.45) -- (167.38,45.77) -- (167.94,49.05) -- (165,47.5) -- (162.06,49.05) -- (162.62,45.77) -- (160.24,43.45) -- (163.53,42.98) -- cycle ;
\draw  [draw opacity=0] (150,40) -- (180,40) -- (180,50) -- (150,50) -- cycle ;

\end{tikzpicture}}) which is a 2$^\text{nd}$ order hyper-edge.
                    Finally, the \textit{SVR Controller} (\resizebox{0.08\linewidth}{!}{\tikzset{every picture/.style={line width=0.75pt}} 

\begin{tikzpicture}[x=0.75pt,y=0.75pt,yscale=-1,xscale=1]

\draw [color={rgb, 255:red, 208; green, 2; blue, 27 }  ,draw opacity=1 ][line width=1.5]    (25,15) -- (35,15) ;
\draw  [draw opacity=0][fill={rgb, 255:red, 0; green, 0; blue, 0 }  ,fill opacity=1 ] (32,15) .. controls (32,13.34) and (33.34,12) .. (35,12) .. controls (36.66,12) and (38,13.34) .. (38,15) .. controls (38,16.66) and (36.66,18) .. (35,18) .. controls (33.34,18) and (32,16.66) .. (32,15) -- cycle ;
\draw  [color={rgb, 255:red, 208; green, 2; blue, 27 }  ,draw opacity=1 ][fill={rgb, 255:red, 255; green, 255; blue, 255 }  ,fill opacity=1 ][line width=1.5]  (25,10) -- (26.47,12.98) -- (29.76,13.45) -- (27.38,15.77) -- (27.94,19.05) -- (25,17.5) -- (22.06,19.05) -- (22.62,15.77) -- (20.24,13.45) -- (23.53,12.98) -- cycle ;
\draw [color={rgb, 255:red, 208; green, 2; blue, 27 }  ,draw opacity=1 ][line width=1.5]    (20,20) -- (30,10) ;
\draw  [draw opacity=0] (10,10) -- (40,10) -- (40,20) -- (10,20) -- cycle ;

\end{tikzpicture}}) is a 1$^\text{st}$ order hyper-edge connected to the \textit{SVR Zone} and \textit{SVR Units} via its \textit{Zone} port, and bears a decision variable $\Delta \mathring{V}$ that updates the regulated bus' voltage setpoint.
                }
                \label{fig:rst_all}
            \end{subfigure}    
            \hfill
            \begin{subfigure}[b]{0.48\linewidth}
                \centering
                \resizebox{0.85\linewidth}{!}{\input{rtc_all.tikz}}
                \caption{
                    \textit{RTC Controller} example.
                    RTCs are TWTs that modify their transformation ratios to regulate the voltage of a bus.
                    In the H2MG representation (right), \textit{TWTs} (\resizebox{0.08\linewidth}{!}{\tikzset{every picture/.style={line width=0.75pt}} 

\begin{tikzpicture}[x=0.75pt,y=0.75pt,yscale=-1,xscale=1]

\draw [line width=1.5]    (115,65) -- (125,61) ;
\draw [line width=1.5]    (116.67,65.67) -- (125,69) ;
\draw [color={rgb, 255:red, 0; green, 0; blue, 0 }  ,draw opacity=1 ][line width=1.5]    (105,65) -- (115,65) ;
\draw  [fill={rgb, 255:red, 255; green, 255; blue, 255 }  ,fill opacity=1 ][line width=1.5]  (113.33,62) .. controls (113.94,62) and (114.51,62.16) .. (115,62.45) .. controls (115.49,62.16) and (116.06,62) .. (116.67,62) .. controls (118.51,62) and (120,63.49) .. (120,65.33) .. controls (120,67.17) and (118.51,68.67) .. (116.67,68.67) .. controls (116.06,68.67) and (115.49,68.5) .. (115,68.22) .. controls (114.51,68.5) and (113.94,68.67) .. (113.33,68.67) .. controls (111.49,68.67) and (110,67.17) .. (110,65.33) .. controls (110,63.49) and (111.49,62) .. (113.33,62) -- cycle ;
\draw  [fill={rgb, 255:red, 255; green, 255; blue, 255 }  ,fill opacity=1 ][line width=1.5]  (110,65.33) .. controls (110,63.49) and (111.49,62) .. (113.33,62) .. controls (115.17,62) and (116.67,63.49) .. (116.67,65.33) .. controls (116.67,67.17) and (115.17,68.67) .. (113.33,68.67) .. controls (111.49,68.67) and (110,67.17) .. (110,65.33) -- cycle ;

\draw  [draw opacity=0][fill={rgb, 255:red, 0; green, 0; blue, 0 }  ,fill opacity=1 ] (122,69) .. controls (122,67.34) and (123.34,66) .. (125,66) .. controls (126.66,66) and (128,67.34) .. (128,69) .. controls (128,70.66) and (126.66,72) .. (125,72) .. controls (123.34,72) and (122,70.66) .. (122,69) -- cycle ;
\draw  [draw opacity=0][fill={rgb, 255:red, 0; green, 0; blue, 0 }  ,fill opacity=1 ] (102,65) .. controls (102,63.34) and (103.34,62) .. (105,62) .. controls (106.66,62) and (108,63.34) .. (108,65) .. controls (108,66.66) and (106.66,68) .. (105,68) .. controls (103.34,68) and (102,66.66) .. (102,65) -- cycle ;
\draw  [draw opacity=0][fill={rgb, 255:red, 0; green, 0; blue, 0 }  ,fill opacity=1 ] (122,61) .. controls (122,59.34) and (123.34,58) .. (125,58) .. controls (126.66,58) and (128,59.34) .. (128,61) .. controls (128,62.66) and (126.66,64) .. (125,64) .. controls (123.34,64) and (122,62.66) .. (122,61) -- cycle ;
\draw  [draw opacity=0] (100,60) -- (130,60) -- (130,70) -- (100,70) -- cycle ;

\end{tikzpicture}}) are 3$^\text{rd}$ order hyper-edges.
                    \textit{TWTs} that are also \textit{RTCs} -- such as the red one in the SLD representation (left) -- are connected via their \textit{TWT} port to \textit{RTCs} (\resizebox{0.08\linewidth}{!}{\tikzset{every picture/.style={line width=0.75pt}} 

\begin{tikzpicture}[x=0.75pt,y=0.75pt,yscale=-1,xscale=1]

\draw [color={rgb, 255:red, 0; green, 0; blue, 0 }  ,draw opacity=1 ][line width=1.5]    (15,15) -- (35,15) ;
\draw  [draw opacity=0][fill={rgb, 255:red, 0; green, 0; blue, 0 }  ,fill opacity=1 ] (32,15) .. controls (32,13.34) and (33.34,12) .. (35,12) .. controls (36.66,12) and (38,13.34) .. (38,15) .. controls (38,16.66) and (36.66,18) .. (35,18) .. controls (33.34,18) and (32,16.66) .. (32,15) -- cycle ;
\draw  [draw opacity=0][fill={rgb, 255:red, 0; green, 0; blue, 0 }  ,fill opacity=1 ] (12,15) .. controls (12,13.34) and (13.34,12) .. (15,12) .. controls (16.66,12) and (18,13.34) .. (18,15) .. controls (18,16.66) and (16.66,18) .. (15,18) .. controls (13.34,18) and (12,16.66) .. (12,15) -- cycle ;
\draw  [draw opacity=0] (10,10) -- (40,10) -- (40,20) -- (10,20) -- cycle ;
\draw  [color={rgb, 255:red, 0; green, 0; blue, 0 }  ,draw opacity=1 ][fill={rgb, 255:red, 255; green, 255; blue, 255 }  ,fill opacity=1 ][line width=1.5]  (20,12) -- (26.67,12) -- (26.67,18.67) -- (20,18.67) -- cycle ;
\draw  [color={rgb, 255:red, 0; green, 0; blue, 0 }  ,draw opacity=1 ][fill={rgb, 255:red, 255; green, 255; blue, 255 }  ,fill opacity=1 ][line width=1.5]  (26.67,12) -- (30,12) -- (30,18.67) -- (26.67,18.67) -- cycle ;

\end{tikzpicture}}), which are 2$^\text{nd}$ order hyper-edges also connected to the regulated bus on the other side.
                    Finally, controllable \textit{RTCs} are connected via their \textit{TWT} port to \textit{RTC Controllers} (\resizebox{0.08\linewidth}{!}{\tikzset{every picture/.style={line width=0.75pt}} 

\begin{tikzpicture}[x=0.75pt,y=0.75pt,yscale=-1,xscale=1]

\draw [color={rgb, 255:red, 208; green, 2; blue, 27 }  ,draw opacity=1 ][line width=1.5]    (25,15) -- (35,15) ;
\draw  [draw opacity=0][fill={rgb, 255:red, 0; green, 0; blue, 0 }  ,fill opacity=1 ] (32,15) .. controls (32,13.34) and (33.34,12) .. (35,12) .. controls (36.66,12) and (38,13.34) .. (38,15) .. controls (38,16.66) and (36.66,18) .. (35,18) .. controls (33.34,18) and (32,16.66) .. (32,15) -- cycle ;
\draw  [draw opacity=0] (10,10) -- (40,10) -- (40,20) -- (10,20) -- cycle ;
\draw  [color={rgb, 255:red, 208; green, 2; blue, 27 }  ,draw opacity=1 ][fill={rgb, 255:red, 255; green, 255; blue, 255 }  ,fill opacity=1 ][line width=1.5]  (20,12) -- (26.67,12) -- (26.67,18.67) -- (20,18.67) -- cycle ;
\draw  [color={rgb, 255:red, 208; green, 2; blue, 27 }  ,draw opacity=1 ][fill={rgb, 255:red, 255; green, 255; blue, 255 }  ,fill opacity=1 ][line width=1.5]  (26.67,12) -- (30,12) -- (30,18.67) -- (26.67,18.67) -- cycle ;

\draw [color={rgb, 255:red, 208; green, 2; blue, 27 }  ,draw opacity=1 ][line width=1.5]    (30,10) -- (20,20) ;

\end{tikzpicture}}) which are 1$^\text{st}$ order hyper-edges that bear a one-hot decision vector $[\mathbb{1}_{0\%}, \mathbb{1}_{2\%}, \mathbb{1}_{5\%}, \mathbb{1}_{7\%}]$ controlling the voltage setpoint of the regulated bus.
                    In general, not all TWTs are RTCs, and not all RTCs are controllable.
                }
                \label{fig:rtc_all}
            \end{subfigure}
            \caption{
                Illustration of the various levers for action.
                On the left part of each subfigure is displayed a \emph{Single Line Diagram} (SLD) of small simplistic systems where controllable devices are shown in red.
                On the right part is displayed the corresponding H2MG representation, where dots (\protect\resizebox{0.02\linewidth}{!}{\protect\tikzset{every picture/.style={line width=0.75pt}} 

\begin{tikzpicture}[x=0.75pt,y=0.75pt,yscale=-1,xscale=1]

\draw  [draw opacity=0][fill={rgb, 255:red, 0; green, 0; blue, 0 }  ,fill opacity=1 ] (12,15) .. controls (12,13.34) and (13.34,12) .. (15,12) .. controls (16.66,12) and (18,13.34) .. (18,15) .. controls (18,16.66) and (16.66,18) .. (15,18) .. controls (13.34,18) and (12,16.66) .. (12,15) -- cycle ;
\draw  [draw opacity=0] (10,10) -- (20,10) -- (20,20) -- (10,20) -- cycle ;

\end{tikzpicture}}) correspond to addresses and all other symbols are hyper-edges defined in Table \protect\ref{tab:classes}.
            }
            \label{fig:levers}
        \end{figure*}

        \paragraph{Context Variable $x$}
        Let $x \in \mathcal{X}$ be a real-life power grid forecast -- referred to as a \emph{context} -- which encompasses both its \textit{structure} and \textit{features}.
        This H2MG is composed of hyper-edges of different classes $c \in \mathcal{C} = \{$\textit{Bus}, \textit{Load}, \textit{Battery}, \textit{Static Var Compensator (SVC)}, \textit{Voltage Source Converter (VSC) Station}, \textit{High Voltage Direct Current (HVDC) Line}, \textit{Line}, \textit{Line Controller}, \textit{Shunt}, \textit{Shunt Controller}, \textit{Generator}, \textit{Secondary Voltage Regulator (SVR) Unit}, \textit{SVR Zone}, \textit{SVR Controller}, \textit{Two Windings Transformer (TWT)}, \textit{Ratio Tap Changer (RTC)}, \textit{RTC Controller}$\}$ (detailed in Table \ref{tab:classes}). 
        Hyper-edges are interconnected via addresses $\mathcal{A}_x \subset \mathbb{N}$ through class-specific ports.

        \paragraph{Decision Variable $y$}
        For a given context $x$, variable $y$ contains decision features for each \textit{Line Controller}, \textit{Shunt Controller}, \textit{SVR Controller} and \textit{RTC Controller} available in $x$.
        It alters the operating condition $x$ as follows.
        \begin{itemize}[noitemsep,topsep=0pt]
            \item \textit{Line Controllers} are associated with a binary decision variable \textit{Connected} that controls the connection status of their associated \textit{Lines} (see Figure \ref{fig:line_all}).
            \item \textit{Shunt Controllers} are associated with a binary decision variable \textit{Switch Status} that controls whether the associated \textit{Shunts} status should be switched (see Figure \ref{fig:shunt_all}).
            \item \textit{SVR Controllers} are associated with a continuous decision variable $\Delta \mathring{V}$ that updates the preexisting regulated bus voltage setpoints $\mathring{V}$ (see Figure \ref{fig:rst_all}).
            \item \textit{RTC Controllers} are associated with a one-hot (\textit{i.e.} categorical) decision vector $[\mathbb{1}_{0\%}, \mathbb{1}_{2\%}, \mathbb{1}_{5\%}, \mathbb{1}_{7\%}]$ defining the regulated bus voltage setpoint as either $100\%$, $102\%$, $105\%$ or $107\%$ of its nominal voltage (see Figure \ref{fig:rtc_all}), which are the only possible values allowed by the voltage management tool used by operators at RTE.
        \end{itemize}

        \paragraph{Objective Function $f$}
        Acknowledging that there is no unique way of defining how good voltages are when applying decision $y$ in context $x$, we choose the following objective,
        \begin{align}
            f(y;x) = f_V(y;x) + f_I(y;x) + f_J(y;x),
            \label{eq:objective_function}
        \end{align}
        where $f_V$ (resp. $f_I$) quantifies voltages (resp. current) violations and $f_J$ quantifies Joule losses.
        Computing $f$ involves the following steps.
        \begin{enumerate}[noitemsep,topsep=0pt]
            \item Load power grid context $x$.
            \item Update the grid according to decision $y$.
            \item Run a static AC simulation.
            \item Extract voltage magnitudes, currents and Joule losses.
        \end{enumerate}
        Then the following quantities are computed,
        \begin{align}
            f_V(y;x) &= \lambda_V \sum_{e \in \mathcal{E}_x^\text{Bus}} \mathbb{1}_{\text{opt},e} \! \times \! \max( 0, \epsilon_V \text{-} v_e, v_e \text{-} 1 \text{+} \epsilon_V )^2, \\
            f_I(y;x) &= \lambda_I \sum_{e \in \mathcal{E}_x^\text{Line} \cup \mathcal{E}_x^\text{TWT}} \mathbb{1}_{\text{opt},e} \! \times \! \max( 0, |i_e| \text{-} 1 \text{+} \epsilon_I )^2, \\
            f_J(y;x) &= \lambda_J \sum_{e \in \mathcal{E}_x^\text{Line} \cup \mathcal{E}_x^\text{TWT}} \mathbb{1}_{\text{opt},e} \! \times \! \vert P_{1,e} + P_{2,e} \vert, \label{eq:joule}
        \end{align}
        where $\forall e \in \mathcal{E}_x^\text{Bus}, v_e = (V_e - \underline{V}_e)/(\overline{V}_e - \underline{V}_e)$ and $\forall e \in \mathcal{E}_x^\text{Line} \cup \mathcal{E}_x^\text{TWT}, i_e = I_e / \overline{I}_e$, $\lambda_V$, $\lambda_I$ and $\lambda_J$ are hyper-parameters, and $\mathbb{1}_{\text{opt},e}$ is equal to 1 if the object should be included in the objective function.
        Notice that the above quantities depend on $(x,y)$, although the dependency has been hidden for the sake of conciseness.
        
        We aim at solving the following black-box optimization problem,
        \begin{align}
            y^\star(x) \in \underset{y}{\arg \min} \ f(y;x).
            \label{eq:initial_optimization_problem}
        \end{align}

    \subsection{Continuous Surrogate Optimization Problem}
        The lack of a proper gradient for objective function $f$ prevents us from directly employing deep learning tools, for they essentially rely on gradient descent.
        Observing that Problem \eqref{eq:initial_optimization_problem} is a single-step \emph{Reinforcement Learning} (RL) problem \cite{sutton}, we propose to transform it as follows.
    
        Let us introduce a continuous surrogate decision variable $z$, along with a probability distribution $\rho$ over the set of decisions $y$.
        We choose to split $z$ and $\rho$ into class-specific components $(y_c)_{c\in \mathcal{C}'}$ and $(\rho^c)_{c \in \mathcal{C}'}$, where $\mathcal{C}' = \{$\textit{Line Controller}, \textit{Shunt Controller}, \textit{SVR Controller}, \textit{RTC Controller}$\}$ is the set of controller classes.
        \begin{align}
            \rho(y|z) = \prod_{c\in \mathcal{C}'} \rho^c(y^c|z^c).
        \end{align}
        Moreover, for a given controllable class $c\in \mathcal{C}'$, the surrogate decision variable $z^c$ is split element-wise,
        \begin{align}
            \forall c \in \mathcal{C}', \rho^c(y^c|z^c) = \prod_{e \in \mathcal{E}^c_x} \rho^c(y_e|z_e). \label{eq:rho_split}
        \end{align}
        In the following, we introduce the definitions of the four conditional probability distributions $(\rho^c)_{c \in \mathcal{C}'}$.

        \paragraph{Line Controllers}
            For each $e \in \mathcal{E}_x^\text{Line Controller}$, the decision is a binary variable $y_e \in \{ 0, 1 \}$.
            Let us introduce a continuous surrogate decision variable $z_e \in \mathbb{R}$ that induces the following conditional distribution,
            \begin{align}
                \rho^\text{Line Controller}(y_e|z_e) = \frac{e^{y_e z_e}}{1 + e^{z_e}}.
            \end{align}
    
        \paragraph{Shunt Controllers}
            Decision variables born by \textit{Shunt Controllers} being also binary variables, we use the exact same approach as for \textit{Line Controllers}.
    
        \paragraph{SVR Controllers}
            For each $e \in \mathcal{E}_x^\text{SVR Controller}$, the decision is a continuous variable $y_e \in \mathbb{R}$.
            Let us introduce a continuous surrogate decision variable $z_e \in \mathbb{R}$ that induces the following conditional Gaussian distribution,
            \begin{align}
                \rho^\text{SVR Controller}(y_e|z_e) = \frac{1}{\sigma \sqrt{2\pi}} e^{ - \frac{( y_e - z_e )^2}{2 \sigma^2}},
            \end{align}
            where $\sigma$ is a fixed hyper-parameter.
    
        \paragraph{RTC Controllers}
            For each $e \in \mathcal{E}_x^\text{RTC Controller}$, the decision variable is a 4-dimensional one-hot vector $y_e \in \{ 0, 1\}^4$ such that $\mathbb{1}^\top . y_e = 1$.
            Let us introduce a continuous surrogate decision variable $z_e \in \mathbb{R}^4$ that associates each category with a score and induces the following conditional distribution,
            \begin{align}
                \rho^\text{RTC Controller}(y_e|z_e) = \frac{e^{y_e^\top . z_e}}{\mathbb{1}^\top . e^{z_e}}.
            \end{align}
    
        Let us introduce $q_\beta(\cdot | x)$ as the Boltzmann distribution derived from $f$ and a temperature parameter $\beta > 0$,
        \begin{align}
            \forall y, q_\beta(y|x) = \frac{e^{-\beta f(y;x)}}{ Z_\beta(x) },
        \end{align}
        where $Z_\beta(x) = \sum_{y'} e^{-\beta f(y';x)}$.
        It associates higher probabilities to decisions $y$ that yield a low objective $f(y;x)$.
    
        Then let us choose as a surrogate objective function $f_\rho^\beta$ the Kullback-Leibler divergence from $\rho(\cdot|z)$ to $q_\beta(\cdot|x)$,
        \begin{align}
            \! \! f^\beta_\rho(z;x) &:= D_{KL}(\rho(\cdot|z) \Vert q_\beta(\cdot|x)), \nonumber\\
                &= -H_\rho(z) + \beta \mathbb{E}_{y \sim \rho(\cdot|z)} \left[ f(y;x) \right] + \log Z_\beta(x). \label{eq:surrogate_objective}
        \end{align}
        where $H_\rho(z)$ is the entropy of $\rho(\cdot|z)$.
        We propose to replace the initial black-box optimization problem \eqref{eq:initial_optimization_problem} with the following continuous surrogate optimization problem,
        \begin{align}
            z_\beta^\star(x) \in \underset{z}{\arg\min}\ f^\beta_\rho(z;x).
            \label{eq:surrogate_optimization_problem}
        \end{align}
        This change of problem is motivated by the intuition that if $z$ is a good solution of problem \eqref{eq:surrogate_optimization_problem}, then $y_\rho(z)\in \arg\max_y \rho(y|z)$ should be a good solution of problem \eqref{eq:initial_optimization_problem}.  Moreover, given the factorization of $\rho(y|z)$, $y_\rho(z)$ may easily be computed component-wise.
    
    \subsection{Surrogate Gradient Estimation}
    
        The surrogate objective function $f_\rho^\beta$ is continuous and differentiable \textit{w.r.t.} $z$, which enables the use of gradient-based methods.
        Using the log-trick ($\nabla_z \phi(z)=\phi(z)\nabla_z\log\phi(z)$), its derivative is expressed as follows,
        \begin{align}
            &g^\beta_\rho(z;x) := \nabla_z f_\rho(z;x), \nonumber \\
            &\quad = -\nabla_z H_\rho(z) + \beta \mathbb{E}_{y \sim \rho(\cdot|z)} \left[ f(y;x) \nabla_z \log \rho(y|z) \right]. \label{eq:surrogate_gradient}
        \end{align}
        Using equation \eqref{eq:rho_split}, $\nabla_z H_\rho(z)$ and $\nabla_z \log \rho(y|z)$ can be split into class and hyper-edge specific terms,
        \begin{align}
            \nabla_z H_\rho(z) &= (\nabla_{z_e}H_{\rho^c}(z_e))_{c \in \mathcal{C}', e \in \mathcal{E}^c_x}, \\
            \nabla_z \log \rho(y|z) &= (\nabla_{z_e}\log \rho^c(y_e|z_e))_{c \in \mathcal{C}', e \in \mathcal{E}^c_x},
        \end{align}
        where all terms are defined in Table \ref{tab:entropy}.

        \begin{table}[h!]
            \centering
            \begingroup
            \renewcommand{\arraystretch}{2.5} 
            \begin{tabular}{lcc}
                Class                   & $\nabla_{z_e}H_{\rho^c}(z_e)$                         & $\nabla_{z_e}\log \rho^c(y_e|z_e)$        \\
                \hline
                Line Controller         & $- z_e \times \dfrac{e^{z_e}}{(1+e^{z_e})^2}$         & $y_e - \dfrac{e^{z_e}}{1+e^{z_e}}$         \\
                Shunt Controller        &  $- z_e \times \dfrac{e^{z_e}}{(1+e^{z_e})^2}$        & $y_e - \dfrac{e^{z_e}}{1+e^{z_e}}$         \\
                SVR Controller          & $0$                                                   & $\dfrac{1}{\sigma^2} (y_e - z_e)$          \\
                RTC Controller          & $- z_e \times \dfrac{e^{z_e} (\mathbb{1}^\top . e^{z_e}-e^{z_e})}{(\mathbb{1}^\top . e^{z_e})^2}$     & $y_e - \dfrac{e^{z_e}}{1+e^{z_e}}$\\
            \end{tabular}
            \endgroup
            \caption{Formulas for derivatives of the entropy and log-probability for the four classes of levers.}
            \label{tab:entropy}
        \end{table}
    
        While the entropy term in equation \eqref{eq:surrogate_gradient} has a closed-form expression, the expectation can only be estimated, e.g. through a  Monte Carlo method, 
        \begin{align}
            \hat{g}^\text{MC}(z;x) =-\nabla_z H_\rho(z) + \frac{\beta}{N} \sum_{i=1}^N f_i \nabla_z\log\rho(y_i|z), \label{eq:MC}
        \end{align}
        where $f_i=f(y_i;x)$, with $(y_i)_{i=1}^N$ i.i.d. samples from $\rho(\cdot | z)$.
        Unfortunately, this raw MC estimator is known to suffer from a large variance and fails to capture a relevant and consistent direction of improvement.
        The following adjustments allow for a more relevant estimator.
        \begin{itemize}[noitemsep,topsep=0pt]
            \item Instead of sampling the decision variables for all classes of objects at the same time, we decompose the gradient estimation by class. During the estimation of a class gradient, we fix all decisions of other class to the most probable decision.
            \item We replace scores $f_i$ from equation \eqref{eq:MC} with clipped scores $f'_i$ defined as:
            \begin{align}
                f'_i = \tanh{\left( \frac{f_i - f(y_\rho(z);x)}{\tau} \right)},
            \end{align}
            where $\tau$ is a hyper-parameter and $y_\rho(z)$ is the most probable decision knowing $z$.
            \item In the case of binary and categorical variables, we sample only unary modifications of $y_\rho(z)$.
        \end{itemize}

    \subsection{Amortized Optimization}
        We aim at solving problem \eqref{eq:surrogate_optimization_problem} not only for a single context $x$, but for a whole distribution $p$ of contexts.
        Thus, we introduce a mapping $\hat{z}_\theta$ -- parameterized by a vector $\theta$ -- that maps contexts $x$ to surrogate decisions $z$.
        This mapping should respect the H2MG structure of the data, and be compatible with variations displayed in the real-life distribution $p$.
    
        We reuse the H2MG-based \emph{Neural Ordinary Differential Equation} (NODE) \cite{ChenRBD18} introduced in previous work \cite{donon2024}, which is a type of GNN especially designed to process H2MGs.
        It relies on the continuous propagation of information between direct neighbors, by associating each address $a\in \mathcal{A}_x$ with a time-varying latent vector $h^t_a\in \mathbb{R}^d$, where $t \in \left[0, 1\right]$ is an artificial time variable and $d\in \mathbb{N}$.
        \begin{align}
            &\!\!\!\forall c \in \mathcal{C}, \forall e\in \mathcal{E}^c_x, \Tilde{x}_e = E_\theta^c(x_e), \label{eq:h2mgnode_encoder}\\
            &\!\!\!\forall a \in \mathcal{A}_x,\! \begin{cases}
                h_a^{t=0} = [0, \dots, 0],  \\
                \! \dfrac{dh_a^t}{dt}\! =\! F_\theta \! \! \left[ h_a^t, \! \tanh\!\left(\sum\limits_{\substack{(c,e,o)\\ \in \mathcal{N}_x(a)}} \! \! M_\theta^{c,o}(h^t_e, \Tilde{x}_e) \!\! \right) \! \! \right]\!\!, \! \!
            \end{cases} \label{eq:h2mgnode_node}\\
            &\!\!\!\forall c \in \mathcal{C}', \forall e\in \mathcal{E}^c_x, [\hat{z}_\theta(x)]_e = D_\theta^c(\Tilde{x}_e, h_e^{t=1}), \label{eq:h2mgnode_decoder}
        \end{align}
        where $\mathcal{N}_x(a) = \{(c,e,o) | e\in \mathcal{E}^c_x, o(e)=a\}$ is the neighborhood of address $a$, $h_e=(h_{o(e)})_{o \in \mathcal{O}^c}$ is the concatenation of the latent vectors of addresses it is connected to and functions $(E_\theta^c)_{c\in \mathcal{C}}$, $F_\theta$, $(M^{c,o}_\theta)_{c\in \mathcal{C}, o \in \mathcal{O}^c}$ and $(D^c_\theta)_{c\in \mathcal{C}'}$ are basic \emph{Multi-Layer Perceptrons} (MLPs).
        
        We now consider the following AO problem,
        \begin{align}
            \theta^\star \in \underset{\theta}{\operatorname{argmin}}\ \mathbb{E}_{x\sim p}\left[ f^\beta_\rho(\hat{z}_\theta(x);x)\right]. \label{eq:amortized_problem}
        \end{align}
        For a given context $x$, the following holds,
        \begin{align}
            \nabla_\theta f^\beta_\rho(\hat{z}_\theta(x); x) = J_\theta \left[ \hat{z}_\theta \right](x)^\top . g^\beta_\rho(\hat{z}_\theta(x); x),
        \end{align}
        where $J_\theta \left[ \hat{z}_\theta \right](x)$ is the Jacobian matrix of the GNN $\hat{z}_\theta$ estimated via automatic differentiation.
        From this basic chain-rule we derive a GNN training loop, shown in Algorithm \ref{alg:training_loop} in the simplifying case of a size 1 minibatch.
        In the actual implementation, multiple contexts are sampled in step (a) and processed in parallel in steps (b) and (c), and the average backpropagated gradient is employed to update $\theta$ in step (d).
        During step (c), the optimization problem \ref{eq:initial_optimization_problem} is not fully solved. Instead, we simply estimate a direction of improvement $\hat{g}$.
        \begin{algorithm}
        \caption{Amortized Optimization Training Loop}\label{alg:training_loop}
        \begin{algorithmic}
        \Require $p$, $\theta$, $\hat{z}_\theta$, $\hat{g}^\beta_\rho$, $\alpha$
        \While{not converged}
            \State $x \sim p$ \Comment{Context sampling (a)}
            \State $\hat{z} \gets \hat{z}_\theta(x)$ \Comment{GNN prediction (b)}
            \State $\hat{g} \gets \hat{g}^\beta_\rho(\hat{z};x)$ \Comment{Gradient estimation (c)}
            \State $\theta \gets \theta - \alpha J_\theta[\hat{z}_\theta](x)^\top . \hat{g}$ \Comment{Back-propagation (d)}
        \EndWhile
        \end{algorithmic}
        \end{algorithm}

\section{Case Study}
\label{sec:case_study}

We aim at providing a post-processing tool over the existing power grid forecasting tool at RTE.
This tool should take as input a real-life power grid forecast -- which usually displays significant voltage violations -- and output targets for the previously introduced levers.
Those targets should be applied over the corresponding forecast, and then a static AC power flow simulation should be run.
The experiment described in the present section is a major step towards this ambitious goal, as training is performed using the real-life and full-scale data available at RTE.

    \subsection{Datasets}

    In this study, we consider the power grid forecasts generated every hour for the next 24 hours for the full French HV-EHV system  with a reduced model of neighboring countries.
    We consider the following datasets.
    \begin{itemize}[noitemsep,topsep=0pt]
        \item \textit{Train} (from Sep. 1$^\text{st}$ to Nov. 30$^\text{th}$ 2024) : 149,249 contexts.
        \item \textit{Val} (from Dec. 1$^\text{st}$ to Dec. 15$^\text{th}$ 2024) : 9,284 contexts.
        \item \textit{Test} (from Dec. 16$^\text{th}$ to Dec. 31$^\text{th}$ 2024) : 9,264 contexts.
    \end{itemize}
    
    Power grid operating conditions (\textit{i.e.} contexts) are framed as H2MGs with at most 24,445 unique addresses and composed of at most 7,643 \textit{Buses}, 8,451 \textit{Loads}, 14 \textit{Batteries}, 7 \textit{SVCs}, 12 \textit{VSC Stations}, 6 \textit{HVDC Lines}, 8,046 \textit{Lines}, 360 \textit{Shunts}, 6,075 \textit{Generators}, 169 \textit{SVR Units}, 27 \textit{SVR Zones}, 2,476 \textit{TWTs} and 1,522 \textit{RTCs}.
    In terms of controllable hyper-edges, there are at most 114 \textit{Line Controllers}, 262 \textit{Shunt Controllers}, 27 \textit{SVR Controllers} and 506 \textit{RTC Controllers}.
    All features listed in Table \ref{tab:classes} are directly extracted from the data using PyPowSybl \cite{pypowsybl}, at the exception of the $\mathbb{1}_\text{opt}$ feature of buses, lines and transformers.
    This feature serves to identify whether an object should be included in the objective function or not.
    Here, it is set to 1 for all (non fictitious) objects within France connected to voltage levels higher than 63kV, and 0 otherwise.

    \subsection{Baseline}

    We compare our methodology to a so-called \textit{Init} baseline, defined as follows.
    \textit{Lines} that are initially closed and that can be opened remain untouched, while already opened lines are removed from the file.
    \textit{Shunts} are not altered from the initial file.
    \textit{SVR} voltage setpoints are set to the regulated bus voltage from the initial file, plus a constant and uniform offset. 
    The latter offset is tuned to minimize the objective function over the \textit{Train} set. 
    The rationale is that we wish our \textit{GNN} to perform better than a uniform and constant modification of setpoints.
    In the initial forecast, \textit{RTC} voltage setpoints take continuous values, which we project into the allowed discrete space $\{1, 1.02, 1.05, 1.07\}$ in per unit.
    Notice that this \textit{Init} baseline is rather simple, as there is currently no other available method for the full-scale system. An ongoing concurrent work aims at developing an expert system heuristics inspired by typical operator routines.
    
    \subsection{Experimental Setup}

        Experiments are conducted using our open-source pipeline \textit{EnerGNN}\footnote{https://github.com/energnn/energnn}, that relies on JAX \cite{jax2018github} and Flax \cite{flax2020github}.
        Input features are normalized using a piecewise linear approximation of the empirical cumulative distribution function, as detailed in the supplementary material of \cite{donon2024}.
        H2MGNODE hyper-parameters have been adjusted by trial-and-error for lack of time to perform a comprehensive hyper-parameter optimization.
        Encoders $(E^c_\theta)_{c\in \mathcal{C}}$ are Multi Layer Perceptrons (MLPs) with 2 hidden layers of sizes $(128, 128)$, an output dimension of size $64$ and Leaky ReLU activation functions.
        Addresses latent vectors are of size $64$.
        System \eqref{eq:h2mgnode_node} is solved by Diffrax \cite{kidger2021on}, using an explicit 1$^\text{st}$-order Euler scheme with $\Delta t=0.005$, and backpropagation is performed using the recursive checkpoint adjoint method.
        Function $F_\theta$ is an MLP without any hidden layer with a Leaky ReLU applied over its output.
        Message passing functions $(M^{c,o}_\theta)_{c\in \mathcal{C}, o \in \mathcal{O}^c}$ and decoders $(D_\theta^c)_{c \in \mathcal{C}'}$ are MLPs with hidden layers of sizes $(128, 128)$ and Leaky ReLU activation functions.
        Back-propagated gradients are processed by Adam \cite{kingma2017} with a learning rate of $1\times 10^{-4}$ and standard parameters.

        Gradient estimation is performed using 8 samples for \textit{Line Controllers} and \textit{RTC Controllers}, and 16 samples for \textit{Shunt Controllers} and \textit{SVR Controllers}, $\beta$ is set to $10^{-4}$ and $\tau$ to $0.1$.
        The objective function $f$ is parameterized by $\lambda_V = \lambda_I = 1$, $\lambda_J = 0.1$ and $\epsilon_I=\epsilon_V = 0.05$.
        The AC simulator used is OpenLoadFlow \cite{powsybl_olf}, with transformer voltage control activated, 100 maximum outer loops, SVR activated ("K\_EQUAL\_PROPORTION") and min (resp. max) target and plausible voltages set to 0 (resp. 3), and standard parameters.
        In the case of a non-convergence, the decision $y$ is associated with a prohibitive cost set to $100$, so as to saturate the $\tanh$ clipping.
        Notice that if the most probable decision $y_\rho(x)$ does not converge, then it is very unlikely that samples will converge, making it impossible to define a direction of improvement.
        In such a case, we choose to return a null gradient.
        In order to increase the probability of having well-defined gradients at the beginning of the training -- where GNN predictions are centered around $0$ -- we add a constant offset to GNN outputs defined as follows.
        \begin{itemize}[noitemsep,topsep=0pt]
            \item For \textit{Line Controllers} and \textit{Shunt Controllers} $e$, $z'_e = z_e-2$, which improves the probability of not disconnecting.
            \item For \textit{SVR Controllers}, $z'_e = z_e+y_e^0$ where $y_e^0$ is the same offset used by the \textit{Init} baseline. $\sigma$ is set to $0.0025$p.u.
            \item For \textit{RTC Controllers}, $z'_e = z_e +2 y_e^0$ where $y_e^0$ is the same one-hot vector used by the \textit{Init} baseline.
        \end{itemize}
        
        Training lasted 10,000 iterations, with minibatches of size 4, which takes 5 days using an NVIDIA A10 GPU for forward and backward GNN computations (steps (b) and (d) from Algorithm \ref{alg:training_loop}) and an AMD EPYC 9554 64-Core Processor for sampling and gradient estimation (steps (a) and (c) from Algorithm \ref{alg:training_loop}).
        Notice that here the full \textit{Train} set has not been explored by the model for lack of time.
        Models are evaluated over the \textit{Validation} set after each epoch and after every 1,000 iterations.

    \subsection{Results}

        As displayed in Table \ref{tab:results}, our \textit{GNN} model reduces the average number of voltage violations from 38.8 to 26.6.
        Notice that the \textit{Init} baseline mostly generates over-voltages (32.6 per context on average) and few under-voltages (6.18 per context on average), while the \textit{GNN} is more balanced with $\sim$13 over-voltages and under-voltages on average per context. This overall improvement of voltage magnitudes is performed without any significant impact on branch overflows and losses caused by Joule's effect on \textit{optimized} branches.
        The average number of branch overflows per context goes from 0.836 to 0.840 (\textit{i.e.} a 0.6\% increase), while the mean Joule losses go from 16.20p.u. to 16.28p.u. (\textit{i.e.} a 0.5\% increase).

        \begin{table}[h!]\renewcommand{\tabcolsep}{1.6mm}
            \centering
            \begin{tabular}{lcc}
                Statistics over the \emph{Test} set of 9,264 contexts                   & \textit{Init}     & \textit{GNN}              \\
                \hline
                Mean \# of Over-Voltages                                         & 32.6             & 13.6                     \\
                Mean \# of Under-Voltages                                        & 6.18             & 13.1                     \\
                Mean \# of Voltage Violations                                    & 38.8             & 26.6                     \\
                Mean \# of Branch Overflows                                             & 0.836            & 0.840                    \\
                Mean Joule Losses (p.u.)                                        & 16.20             & 16.28                     \\ 
                \hline
                Mean \% of Disconnected Lines                                   & 0.00\%              & 0.845\%                 \\
                Mean \% of Connected Shunt Capacitors                           & 3.81\%          & 3.90\%                  \\
                Mean \% of Connected Shunt Inductors                            & 74.6\%          & 86.1\%                  \\
                Mean of SVR Voltage Setpoints (p.u.)                            & 1.02             & 1.02                     \\
                Std. dev. of SVR Voltage Setpoints (p.u.)                       & 0.0281           & 0.0265                   \\
                Mean \% of RTCs at 100\%                                         & 67.5\%          & 68.3\%                   \\
                Mean \% of RTCs at 102\%                                         & 25.9\%          & 31.7\%                  \\
                Mean \% of RTCs at 105\%                                         & 5.80\%          & 0.00\%                       \\
                Mean \% of RTCs at 107\%                                         & 0.0835\%        & 0.00\%                       \\
            \end{tabular}
            \caption{
                Metrics and levers usage before and after the action of the trained \textit{GNN} model, over the \textit{Test} set.
            }
            \label{tab:results}
        \end{table}

\begin{figure}[h!]
            \centering
            \begin{subfigure}[b]{0.48\linewidth}
                \centering
                \includegraphics[width = \linewidth]{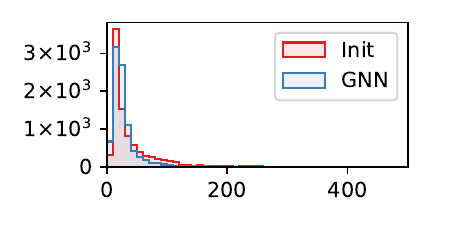}
                \vspace{-2\baselineskip}
                \caption{Linear scale for y-axis.}
                \label{fig:overvoltage_count}
            \end{subfigure}
            \hfill
            \begin{subfigure}[b]{0.48\linewidth}
                \centering
                \includegraphics[width = \linewidth]{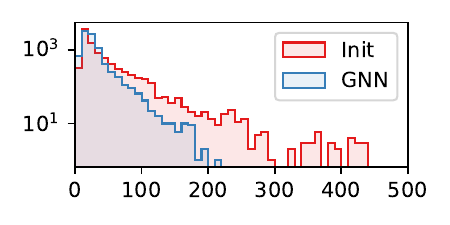}
                \vspace{-2\baselineskip}
                \caption{Logarithmic scale for y-axis.}
                \label{fig:overvoltage_count_log}
            \end{subfigure}
            \caption{
                Histogram of voltage violation counts per context, over the \textit{Test} set. 
                The red curve corresponds to initial values, and the blue curve to the outcome of the GNN's decision.
                (a) and (b) display the same data on different y-axis scales.
            \label{fig:voltage_violation}}
            
    
            \centering
            \includegraphics[width = \linewidth]{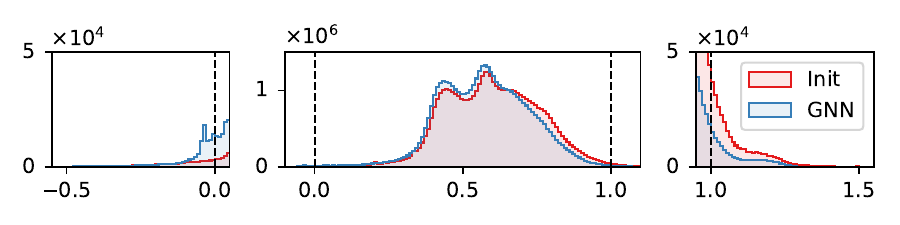}
            \vspace{-1\baselineskip}
            \caption{
                Histogram of normalized voltages over all buses and all contexts from the \textit{Test} set, $v_e=(V_e-\underline{V}_e)/(\overline{V}_e-\underline{V}_e)$. 
                 Values out of the $\left[0, 1\right]$ range correspond to voltage violations.
                The left and right panels focus only on the distribution tails, and share the same scale.
                The y-axis scale is linear.
                \label{fig:voltages}}     
        \end{figure}

    Figure \ref{fig:voltage_violation} shows that the \textit{Init} baseline yields up to $\sim$450 voltage violations per context, while the \textit{GNN} yields at most $\sim$200 voltage violations.
        The slope in logarithmic scale indicates that the \textit{GNN} does a better overall job at reducing the number of violations.
        Figure \ref{fig:voltages} displays the distribution of normalized voltages ($\forall e \in \mathcal{E}^\text{Bus}, v_e=(V_e-\underline{V}_e)/(\overline{V}_e-\underline{V}_e)$) across all optimized buses from the whole \textit{Test} set.
        The trained \textit{GNN} managed to reduce the upper tail of over voltages (where $v_e >1$), at the cost of an increase of the lower tail (where $v_e<0$).
        A deeper analysis of the results shows that most over voltages are on the 225kV and 63kV voltage levels, while the under voltages are mostly on the 400kV voltage level. 


\begin{figure}[h!]
            \centering
            \begin{subfigure}[b]{0.48\linewidth}
                \centering
                \includegraphics[width = \linewidth]{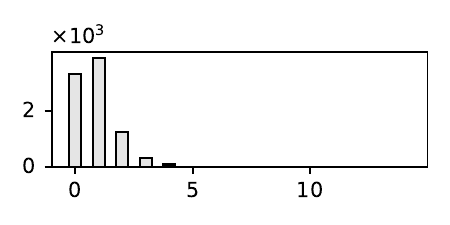}
                \vspace{-2\baselineskip}
                \caption{Number of lines opened by the GNN per context.}
                \label{fig:line_count}
            \end{subfigure}
            \hfill
            \begin{subfigure}[b]{0.48\linewidth}
                \centering
                \includegraphics[width = \linewidth]{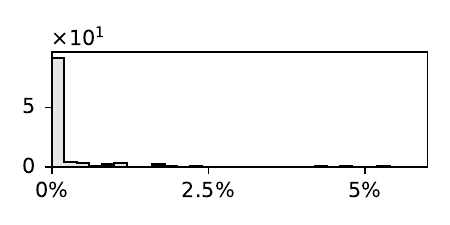}
                \vspace{-2\baselineskip}
                \caption{Percentage of opening by the GNN per line.}
                \label{fig:line_usage}
            \end{subfigure}
            \vfill
            \begin{subfigure}[b]{0.48\linewidth}
                \centering
                \includegraphics[width = \linewidth]{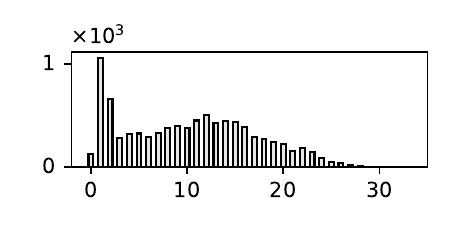}
                \vspace{-2\baselineskip}
                \caption{Number of shunts connected or disconnected by the GNN per context.}
                \label{fig:mcs_count}
            \end{subfigure}
            \hfill
            \begin{subfigure}[b]{0.48\linewidth}
                \centering
                \includegraphics[width = \linewidth]{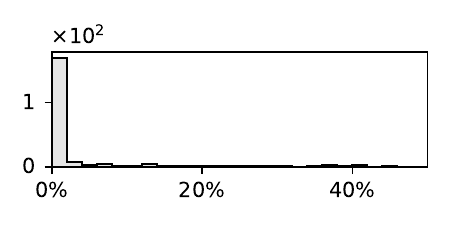}
                \vspace{-2\baselineskip}
                \caption{Percentage of connection or disconnection by the GNN per shunts.}
                \label{fig:mcs_usage}
            \end{subfigure}
            \vfill
            \begin{subfigure}[b]{0.48\linewidth}
                \centering
                \includegraphics[width = \linewidth]{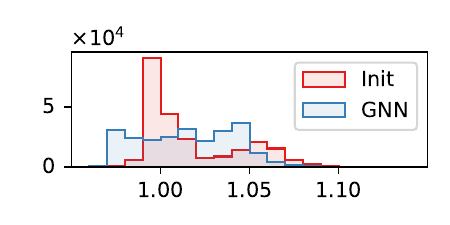}
                \vspace{-2\baselineskip}
                \caption{SVR voltage setpoints.}
                \label{fig:rst_setpoint}
            \end{subfigure}
            \hfill
            \begin{subfigure}[b]{0.48\linewidth}
                \centering
                \includegraphics[width = \linewidth]{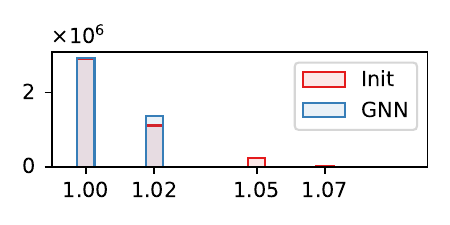}
                \vspace{-2\baselineskip}
                \caption{RTC voltage setpoints.}
                \label{fig:rtc_setpoint}
            \end{subfigure}
            
            \caption{
                Distributions of control levers' setpoints and usages.
                All y-axes are in logarithmic scale, so as to highlight outliers.
            }
            \label{fig:levers}
        \end{figure}

        The \textit{GNN} model exploits the different control levers at its disposal in the following way.
        \begin{itemize}[noitemsep,topsep=0pt]
            \item \textit{Line openings.} The \textit{GNN} very rarely opens transmission lines (0.845\% on average), although Figure \ref{fig:line_count} shows that up to 12 lines can be opened by the \textit{GNN} at the same time. Figure \ref{fig:line_usage} shows that most lines are barely used, while some are opened in more that 5\% of contexts.
            \item \textit{Shunts state switching.} The \textit{GNN} almost never touches the shunt capacitors -- which increase voltages -- ($\sim$3.8\% of them are connected), while the percentage of connected shunt inductors -- which decrease voltages -- goes from 74.6\% to 86.1\%. Some shunts are more used than others (see Figure \ref{fig:mcs_usage}): most of them are barely requested to change states, while some are switched in 45\% of contexts from the \textit{Test} set.
            \item \textit{SVR voltage setpoints.} Figure \ref{fig:rst_setpoint} shows that the \textit{Init} baseline has two main modes, one around $1$p.u. and the other around $1.05$p.u., while the \textit{GNN} seems to exploit a wider range of values, mostly spread between $0.97$p.u. and $1.05$p.u. Despite this difference, their means and standard deviations remain roughly the same (respectively $\sim$$1.015$p.u. and $\sim$$0.027$p.u.).
            \item \textit{RTC voltage setpoints.} The \textit{GNN} model never uses the 105\% and 107\% setpoints, and distributes its setpoints between 100\% (68.32\%) and 102\% (31.68\%).
        \end{itemize}

        Notice that among the 9,264 contexts from the \textit{Test} set, the \textit{Init} baseline makes the AC simulator converge in 98.7\% of them, while the \textit{GNN} model makes it converge in 96.5\% of them.
        The convergence failures occur despite the use of OpenLoadFlow \cite{powsybl_olf}, which is an industrial simulator that implements many robustifying heuristics.
        All these non-converging contexts were excluded from the above analysis.
        At inference time, importing 4 power grid files, applying the GNN model in parallel, modifying the files, running an AC simulation and saving the updated files takes 12s using an A10 GPU and an AMD EPYC 9554 64-Core Processor.

\section{Conclusion}
\label{sec:conclusion}

The work presented in the present study focuses on tertiary voltage control, with the specific aim of improving, in terms of reactive power management, the forecasting pipeline used at RTE in the context of operation planning.
The objective is to reduce the amount of voltage violations in a given forecast by acting on continuous (SVR setpoints), binary (shunts and lines) and categorical (RTC setpoints) control levers.

In the paper, we have explained in details how the H2MG formalism may be used in order to represent in a faithful fashion real-life power system operating conditions and control problems. We describe a methodology that leverages a GNN using the H2MG-based neural ordinary differential equation model (H2MGNODE), trained in a self-supervised mode so as to minimize an objective function evaluated by a (black-box) simulator of the power system physics.
We have experimentally validated this approach on a real-life and full-scale use case, successfully decreasing the number of voltage violations from 38.8 to 26.6 on average, as compared to a simple baseline.
These quite realistic experiments  demonstrate the feasibility of an AI-based decision support tool to assist power system operators in the real world and in real-time.

However, multiple issues arose during the experiments.
First of all, the overall training is currently too slow to allow for a proper hyper-parameter optimization.
This problem could be addressed by accelerating both the simulator and the GNN core, or by increasing the parallelism by using more CPUs and GPUs. More broadly, any notable progress in the context of machine learning from a combination of observational data and simulations shall be scrutinized and potentially leveraged to speed-up the training stage.

Secondly, the static simulator based on AC power flow computations used in the training loop failed to converge in a significant number of contexts encountered during our experiments, despite the many robustifying heuristics implemented.
Considered operating conditions push the simulator far off its nominal operating domain.
While this slowed down the training / validation loops, it did, however, not prevent the approach to learn useful TVC policies. In any case, the self-supervised training loop is designed so as to be able to take advantage of any future progress in terms of power system simulators in a seamless way. Clearly, working on the robustness and accuracy of digital twins of power systems is needed to support the operators to manage operating contexts that will be more and more often atypical. Our work on AI-based decision support tools presented in this paper is in perfect synergy with that other line of work.


\bibliography{biblio}

@article{gnn_original,
  author={Scarselli, Franco and Gori, Marco and Tsoi, Ah Chung and Hagenbuchner, Markus and Monfardini, Gabriele},
  title={{The Graph Neural Network Model}},
  journal={IEEE transactions on neural networks},
  year={2008},
}

@inproceedings{thomaskipfgnn,
  title={{Modeling Relational Data with Graph Convolutional Networks}},
  author={Schlichtkrull, Michael and Kipf, Thomas N and Bloem, Peter and Van Den Berg, Rianne and Titov, Ivan and Welling, Max},
  booktitle={European semantic web conference},
  year={2018},
  organization={Springer}
}

@misc{RTE7k,
    author = {RTE},
    title = {RTE7000},
    URL = {https://huggingface.co/datasets/OpenSynth/RTE700},
    year = {2021}
}

@misc{amortizedoptimization,
      title={{Tutorial on Amortized Optimization}}, 
      author={Brandon Amos},
      howpublished="arXiv: \href{www.arxiv.org/abs/2202.00665}{2202.00665 [cs.LG]}",
      year={2022},
      eprint={2202.00665},
      archivePrefix={arXiv},
      primaryClass={cs.LG},
}

@article{donon2024,
    author = {Donon, Balthazar and Cubélier, François and Karangelos, Efthymios and Wehenkel, Louis and Crochepierre, Laure and Pache, Camille and Saludjian, Lucas and Panciatici, Patrick},
    title = {{Topology-Aware Reinforcement Learning for Tertiary Voltage Control}},
    journal = {Electric Power Systems Research},
    year = {2024},
}

@software{jax2018github,
  author={James Bradbury and Roy Frostig and Peter Hawkins and Matthew James Johnson and Chris Leary and Dougal Maclaurin and George Necula and Adam Paszke and Jake Vander{P}las and Skye Wanderman-{M}ilne and Qiao Zhang},
  title={{JAX}: composable transformations of {P}ython+{N}um{P}y},
  year={2018},
}

@software{flax2020github,
  author={Jonathan Heek and Anselm Levskaya and Avital Oliver and Marvin Ritter and Bertrand Rondepierre and Andreas Steiner and Marc van {Z}ee},
  title={{F}lax: A neural network library and ecosystem for {JAX}},
  year={2024},
}

@book{goodfellow2016deep,
  title={{Deep Learning}},
  author={Goodfellow, Ian and Bengio, Yoshua and Courville, Aaron and Bengio, Yoshua},
  year={2016},
  edition={1},
  publisher={MIT Press Cambridge}
}

@article{hornik1991approximation,
  title={{Approximation Capabilities of Multilayer Feedforward Networks}},
  author={Hornik, Kurt},
  journal={Neural networks},
  year={1991},
  publisher={Elsevier}
}

@phdthesis{kidger2021on,
    title={{O}n {N}eural {D}ifferential {E}quations},
    author={Patrick Kidger},
    year={2021},
    school={University of Oxford},
}

@conference{kingma2017,
    title={{Adam: A Method for Stochastic Optimization}}, 
    author={Diederik P. Kingma and Jimmy Ba},
    booktitle={International Conference for Learning Representations (ICLR)},
    year={2015},
}

@phdthesis{donon2022,
  TITLE = {{Deep Statistical Solvers \& Power Systems Applications}},
  AUTHOR = {Donon, Balthazar},
  SCHOOL = {{Universit{\'e} Paris-Saclay}},
  YEAR = {2022},
}

@INPROCEEDINGS{Fukuyama,
  author={Fukuyama, Y. and Yoshida, H.},
  booktitle={Congress on Evolutionary Computation}, 
  title={{A Particle Swarm Optimization for Reactive Power and Voltage Control in Electric Power Systems}}, 
  year={2001},
}

@book{sutton,
  author = {Sutton, Richard S. and Barto, Andrew G.},
  publisher = {The MIT Press},
  edition = {2},
  title = {{Reinforcement Learning: An Introduction}},
  year = {2018}
}

@misc{liao2021reviewgraphneuralnetworks,
      title={{A Review of Graph Neural Networks and Their Applications in Power Systems}}, 
      author={Wenlong Liao and Birgitte Bak-Jensen and Jayakrishnan Radhakrishna Pillai and Yuelong Wang and Yusen Wang},
      howpublished="arXiv: \href{www.arxiv.org/abs/2101.10025}{2101.10025 [cs.LG]}",
      year={2021},
      eprint={2101.10025},
      archivePrefix={arXiv},
      primaryClass={cs.LG},
}

@misc{pan2022,
      title={{DeepOPF: A Feasibility-Optimized Deep Neural Network Approach for AC Optimal Power Flow Problems}}, 
      author={Xiang Pan and Minghua Chen and Tianyu Zhao and Steven H. Low},
      howpublished="arXiv: \href{www.arxiv.org/abs/2007.01002}{2007.01002 [eess.SY]}",
      year={2022},
      eprint={2007.01002},
      archivePrefix={arXiv},
      primaryClass={eess.SY},
}

@INPROCEEDINGS{9320077,
  author={Thayer, Brandon L. and Overbye, Thomas J.},
  booktitle={IEEE Electric Power and Energy Conferenced}, 
  title={{Deep Reinforcement Learning for Electric Transmission Voltage Control}}, 
  year={2020},
}

@misc{hagmar2022,
      title={{Deep Reinforcement Learning for Long-Term Voltage Stability Control}}, 
      author={Hannes Hagmar and Le Anh Tuan and Robert Eriksson},
      howpublished="arXiv: \href{www.arxiv.org/abs/2207.04240}{2207.04240 [eess.SY]}",
      year={2022},
      eprint={2207.04240},
      archivePrefix={arXiv},
      primaryClass={eess.SY},
}

@article{ZHEN202243,
title = {{Design and Tests of Reinforcement-Learning-Based Optimal Power Flow Solution Generator}},
journal = {Energy Reports},
year = {2022},
author = {Hongyue Zhen and Hefeng Zhai and Weizhe Ma and Ligang Zhao and Yixuan Weng and Yuan Xu and Jun Shi and Xiaofeng He},
}

@misc{lopezcardona2025,
      title={{Proximal Policy Optimization with Graph Neural Networks for Optimal Power Flow}}, 
      author={Ángela López-Cardona and Guillermo Bernárdez and Pere Barlet-Ros and Albert Cabellos-Aparicio},
      howpublished="arXiv: \href{www.arxiv.org/abs/2212.12470}{2212.12470 [cs.AI]}",
      year={2025},
      eprint={2212.12470},
      archivePrefix={arXiv},
      primaryClass={cs.AI},
}

@misc{owerko2022,
      title={{Unsupervised Optimal Power Flow Using Graph Neural Networks}}, 
      author={Damian Owerko and Fernando Gama and Alejandro Ribeiro},
      howpublished="arXiv: \href{www.arxiv.org/abs/2210.09277}{2210.09277 [eess.SY]}",
      year={2022},
      eprint={2210.09277},
      archivePrefix={arXiv},
      primaryClass={eess.SY},
}

@misc{li2022,
      title={{Deep Reinforcement Learning for Optimal Power Flow with Renewables Using Graph Information}}, 
      author={Jinhao Li and Ruichang Zhang and Hao Wang and Zhi Liu and Hongyang Lai and Yanru Zhang},
      howpublished="arXiv: \href{www.arxiv.org/abs/2112.11461}{2112.11461 [cs.LG]}",
      year={2022},
      eprint={2112.11461},
      archivePrefix={arXiv},
      primaryClass={cs.LG},
}

@inproceedings{diehl2019,
  title={{Warm-Starting AC Optimal Power Flow with Graph Neural Networks}},
  author={Diehl, Frederik},
  booktitle={NeurIPS Workshop on Tackling Climate Change with Machine Learning},
  year={2019}
}

@phdthesis{Castillo2016,
    author = {Andrea R. Castillo},
    title = {{Essays on the ACOPF Problem: Formulations, Approximations, and Applications in the Electricity Markets}},
    school = {Johns Hopkins University},
    year = {2016}
}

@inproceedings{ChenRBD18,
  author = {Chen, Tian Qi and Rubanova, Yulia and Bettencourt, Jesse and Duvenaud, David},
  booktitle = {NeurIPS},
  title = {{Neural Ordinary Differential Equations}},
  year = 2018
}

@software{pypowsybl,
  author = {{Committers of PyPowSyBl}},
  title = {{PyPowSyBl, a Python API for PowSyBl Toolbox}},
  url = {http://github.com/powsybl/pypowsybl},
}

@software{powsybl_olf,
  author = {{Committers of OpenLoadFlow}},
  title = {{OpenLoadFlow, a Loadflow for PowSyBl Toolbox}},
  url = {https://github.com/powsybl/powsybl-open-loadflow},
}

@software{powsybl,
  author = {{Committers of PowSyBl}},
  title = {{PowSyBl (Power System Blocks), a Power System Toolbox}},
  url = {https://github.com/powsybl/},
}
\end{document}